\documentclass[draftclsnofoot,onecolumn,12pt]{IEEEtran}
\usepackage{color}
\usepackage{amssymb}

\newtheorem{remark}{Remark}
\newtheorem{lemma}{Lemma}
\newtheorem{proposition}{Proposition}

\usepackage{cite}
\usepackage{caption}
\usepackage{graphicx}
\usepackage{amsmath}
\usepackage{epstopdf}
\usepackage{bm}
\usepackage{supertabular}
\usepackage{multirow}
\newlength\savewidth

\usepackage{stfloats}
\usepackage{color}
\usepackage{algorithm}
\usepackage{algorithmic}
\ifCLASSOPTIONcompsoc
\else
  \usepackage[caption=false,font=footnotesize]{subfig}
\fi
\usepackage{setspace}
\usepackage{multirow}
\usepackage{amsfonts}

\begin{document}
\pagestyle{empty}
%
\title{\LARGE Secure Transmission Design for Virtual Antenna Array-aided  Device-to-device Multicast Communications}

\author{Xinyue Hu, Yibo Yi,\IEEEmembership{Student Member,~IEEE,} Kun Li, Hongwei Zhang and Caihong Kai,\IEEEmembership{ Member,~IEEE}
\thanks{This work was supported by the National Natural Science Foundation of China under Grant 61971176.(Corresponding author: Caihong Kai.)}
\thanks{Xinyue Hu, Yibo Yi and Caihong Kai are with School of Computer Science and Information Engineering, Hefei University of Technology, Hefei, China (e-mail: $\rm {\rm{\{ }}$huxinyue, yiyibo2018${\rm{\} }}$@mail.hfut.edu.cn; $\rm {\rm{\{ }}$chkai${\rm{\} }}$@hfut.edu.cn.). Xinyue Hu is also with School of Electronic and Information Engineering, Anhui University, Hefei, China (e-mail: 21058@ahu.edu.cn.) }
\thanks{Kun Li and Hongwei Zhang are with School of Electronic and Information Engineering, Anhui University, Hefei, China (e-mail: $\rm {\rm{\{ }}$lik, hwzhang${\rm{\} }}$@ahu.edu.cn.).}
}
\maketitle
\begin{abstract}
This paper investigates the physical-layer security in a Virtual Antenna Array (VAA)-aided Device-to-Device Multicast (D2DM) communication system, where the User Equipments (UEs) who cache the common content will form a VAA system and cooperatively multicast the content to UEs who desire it. Specifically, with the target of securing the VAA-aided D2DM communication under the threat of multiple eavesdroppers, we propose a secure beamforming scheme by jointly considering the formed VAA and the Base Station (BS). For obtaining the optimal beamforing vectors, a nonsmooth and nonconvex Weight Sum Rate Maximization Problem (WRMP) is formulated and solved using Successive Convex Approximation (SCA) approach. Furthermore, we consider the worst case that eavesdroppers cooperatively form a eavesdrop VAA to enhance the overhearing capacity. In this case, we modify the securing
beamforming scheme, formulate the corresponding WRMP and solve it using a two-level optimization. Simulation results validate the improvements of the VAA-aided D2DM scheme in terms of communication security compared with conventional D2DM schemes.
\end{abstract}

%
\IEEEpeerreviewmaketitle

\section{Introduction}

\IEEEPARstart{D}{evice}-to-Device Multicast (D2DM) communication which developed from Device-to-Device (D2D) unicast \cite{8245811} communication has been regarded as one of the key technologies to improve the data rate and alleviate the workload of cellular networks \cite{9177351 ,9351975}. By allowing one transmitted information bearing signal to be received and decoded by multiple receivers, D2DM could achieve the large-scale parallel implementation of local contents sharing in high-density environments, thus can support massive connectivity and also reduce the redundancy transmission of the Base Station (BS), signaling overhead, and bandwidth waste. Due to the high spectrum utilization and transmission efficiency, D2DM is not only suitable for the data-rate-hungry mobile applications such as online conference, video streaming, mobile computing, and virtual reality, but also considered as a key enabling technique for various future cellular network services such as mobile edge computing (MEC)\cite{9372880,9488278,8896954,8676385,8502937} and fog-radio access network (F-RAN) \cite{8944005,9512273}.

Although D2DM has a wide range of applications and there have been many literatures on how to improve its network delivery efficiency\cite{9339841,8394984,8882292,9351975,8260862,7892899}, on the other hand, as in other wireless communication systems, communication security is an important indicator to evaluate the performance of a D2DM system. Physical Layer Security (PLS), which is an essential strategy to protect the information by fully utilizing the basic characteristics of the wireless channel and noise, has been regarded as a highly promising security approach for wireless communications. Compared with the cryptography in the higher layer, PLS is independent of the computational complexity and thus more suitable for computation resource limited networks such as D2D communications. Motivated by this, many research efforts have been devoted to secure D2D communication from the physical layer security perspective \cite{6626307,7467465,8332096,7572117,8335294,8438892,9760142,9638994}.  Ref. \cite{6626307} attempted to increase the D2D link's capacity while guaranteeing the security rate by controlling the transmit power of D2D pairs. Refs. \cite{7467465,8332096} considered that multiple D2D pairs can access the same resource block and one D2D pair is also permitted to access multiple RBs. In \cite{7572117}, aimed to secure CUE's communication and increase energy efficiency of D2D communication, the condition whether D2D link could reuse a CUE's RB was derived. Ref. \cite{8335294} investigated both single-channel and multi-channel D2D communications. For providing security to CUEs and improving the spectral efficiency of the D2D links simultaneously, the authors jointly optimized the power allocation and channel assignment of the D2D links and CUEs. In \cite{8438892}, the model selection of a D2D link under the physical-layer security perspective was investigated. The PLS in UAV-enabled D2D communications was investigated in \cite{9760142} where a UAV plays the role of flying BS. In \cite{9638994}, authors shown that the Non-Orthogonal Multiple Access (NOMA) technique could also be used to secure D2D communications. Note that, the above literatures are mainly focused on the D2D unicast scenarios. To the best of our knowledge, the PLS schemes in D2DM are rare \cite{8433158}. No matter whether the design target is to increase network capacity or to provide PLS, the main difference between D2D unicast and D2DM is that the data rate in D2DM is limited by the worst Signal to Interference Plus Noise Ratio (SINR) UE and hence leading to a max-min optimization problem which is hard to deal with.

However, the above secure schemes in D2D communication are mainly based on cooperative jamming, in which the cellular communication signals and other D2D communication signals are utilized to interfere eavesdroppers for securing intended D2D communications, or vice versa. Constrained by the limited physical size, UEs usually have only one antenna or multiple antennas that are less distant than the coherent distance. Hence, the multiple-antenna based beamforming technologies for providing PLS \cite{9512271,9492144,7018097} are not suitable for D2D communications. Thus, how to secure D2D communications is still an open issue and this paper makes attempt to deal with it by introducing the following Virtual Antenna Array (VAA) technique.

Recently, a novel implementation of Multiple-Input-Multiple-Output (MIMO) wireless system called VAA \cite{phd, patent,4161912} was proposed as a promising approach to further increase the PLS performance and network delivery efficiency of the above D2DM scenario. The VAA system is realized by permitting adjacent single antenna mobile terminals to cooperate among each other and thus form the so called virtual antenna array. In the VAA system, as a virtual transmitter, the elements in VAA could perform cooperative beamforming to transmit information to receivers, while as a virtual receiver, the elements in VAA could exchange the received signals via some short-range communication techniques such as blue tooth and then perform some multiple-antenna receive techniques like maximum ratio combination to increase SINR. It is important to note that, compared with the Single-Input-Single-Output (SISO) channel, the spatial diversity of VAA together with cooperative beamforming can significantly improve the quality of the transmitted signal and then enhance the PLS performance\cite{8288649, 9298942} and channel capacity \cite{4161912, 7302604, 5073429,6136816,6560494}.

In this papar, we propose a VAA-aided D2DM (VD2DM) scheme for securing D2DM communication. In the VD2DM, UEs who cache the common contents cooperatively form a VAA system and each UE could be treated as one of the antennas in the VAA. By multiplying the signals each UE sends itself by a beamforming coefficient (i.e., the corresponding element in the beamforming vector), VAA could jointly conduct cooperative beamforming to multicast the common contents to receivers while prevent eavesdroppers from overhearing. Further, unlike the traditional D2D communications which only allow one UE as the transmitter, the VD2DM could take full use of the same the common contents cached on different UEs. The main contributions of this paper are summarized as follows:

\begin{itemize}
\item[1)] We propose a novel secure VD2DM transmission scheme, which integrates the VAA into the D2DM underlying cellular network. In the proposed scheme, all UEs are equipped with one antenna, and the BS is equipped with multiple antennas. A group of UEs (i.e., D2D receivers) needs a common confidential information data and in the D2D transmitter group, each UE (i.e., D2D transmitters) has cached the common confidential information desired by the D2D receivers. Under the coordination of the BS, D2D transmitters form a VAA, where each UE multiplies the signals by a complex beamforming coefficient\footnote{Amplitude control can be accomplished by power control and phase control can be accomplished by delay.}to perform cooperative beamforming for multicasting the confidential information to D2D receivers. Meanwhile, a group of baleful UEs (i.e., eavesdroppers) desire to overhear the information delivered by the VAA. Since the VAA transmission reuses the downlink subcarrier resources of a CUE and hence the downlink signal transmitted by the BS could be treated as a cooperative jamming to secure VD2DM. To further limit the overhearing capacity of eavesdroppers, the BS splits a part of power to transmit an Artificial Noise (AN) to block eavesdroppers.

\item[2)] We explore the difference between the VD2DM and the traditional D2D unicast, and then formulate the weighted sum rate maximization problem (WRMP) to increase the achievable rate of CUE and the achievable secrecy rate of VD2DM simultaneously by jointly optimizing the transmit beamforming of the BS, the cooperative beamforming vector of the VAA and the covariance matrix of AN. The formulated problem is non-convex and non-smooth since both max-min and min-max problem are included, i.e., the achievable performance of the multicast group is generally limited by the receiver with the lowest SINR and the secrecy performance is limited by the eavesdropper with the highest SINR. We then introduce a series of auxiliary variables to transfer the non-convex objective function and constraints into a series of convex objective function and constraints. As last, an effective iteration algorithm is designed to solve it.

\item[3)] Moving beyond that, we investigate an extreme case (i.e., the worst case) in which the eavesdroppers also form a VAA and gang up together to jointly overhearing the communication of VD2DM. In our analysis, we treat eavesdroppers as one equivalent one multi-antenna eavesdropper called Eve. In this case, with the help of semidefinite relaxation and slack operation, we can decompose the WRMP into a two-level optimization problem, in which the outer level is a single-variable optimization, and the inner level is a non-convex problem, which can be recast to a SemiDefinite Problem (SDP) based on Arithmetic Geometry Mean (AGM) inequality.
\end{itemize}

The remainder of the paper is organized as follows. Section II introduces our considered system model. Section III gives our formulated WRMP problem and the detailed solution to the formulated problem. Section IV extends the considered scenario to the worst case and gives the corresponding solution. Simulation results are shown in Section V and finally Section VI concludes the paper.

$Notations$: $\mathbb{C}$ and $\mathbb{R}$ denote the sets of complex and
 real numbers, respectively; $\rm Rank( \bullet )$, $\rm Tr(\bullet)$, ${\left[ {\bullet} \right]^{ - 1}}$, ${\left[ {\bullet} \right]^{ T}}$, and ${\left[ {\bullet} \right]^{ H}}$ denote the rank, trace,
inverse, transpose, and conjugate-transpose operations,
respectively; ${\left\|  \bullet  \right\|_2}$ and ${\left[ { \bullet , \bullet } \right]^ + }$ denote the Frobenius norm and the maximum value of its argument, respectively.
Except for $\bf e$, bold symbols in capital
letter and small letter denote matrices and vectors, respectively and $\bf e$ is the natural constant.
$\bf {A} \succeq \bf {0}$ means $\bf {A}$ is positive semidefinite matrix.
${\bf{a}}[n]$ and ${\bf{A}}[n]$ are the $n^{th}$ element of the vector ${\bf{a}}$ and the $n^{th}$ element of the diagonal entry of the matrix ${\bf{A}}$, respectively.
$\log (\bullet)$ denotes the natural logarithm of its argument; ${{\cal C}{\cal N}}\left( {a,b} \right)$
denotes the complex Gaussian distribution with
mean $a$ and variance $b$; ${{\bf{I}}_L}$ is an $L \times L$ identity matrix.

\section{ System Model and Preliminary}

\begin{figure}[t]
  \centering
  \includegraphics[width=2.8in]{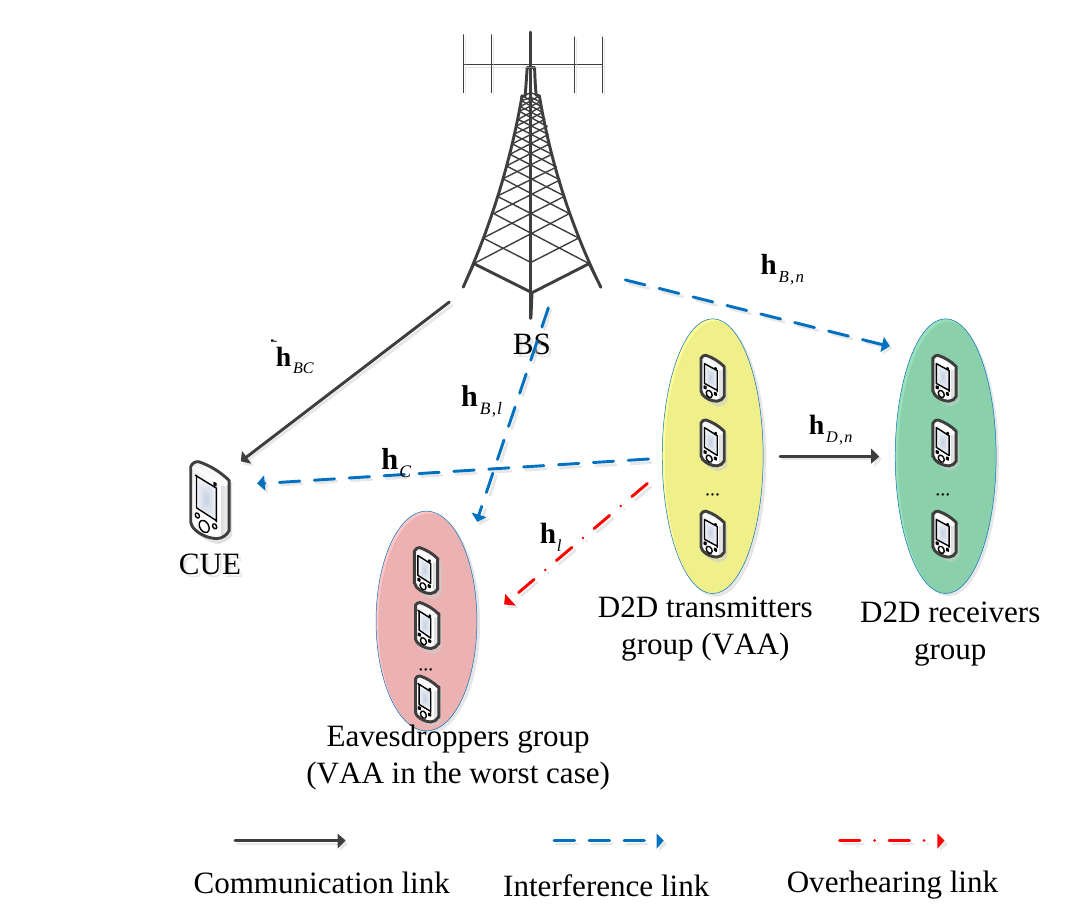}\\
  \caption{System scenario of the considered VD2DM, in which exists $M$ D2D transmitters, $N$ D2D receivers and $L$ eavesdroppers.}
\end{figure}
Consider a VD2DM communication scenario which underlays over the downlink spectrum of a cellular communication network as shown in Fig. 1. A group of D2D receivers want to obtain identical confidential information, and a group of UEs who pre-cache the identical confidential information will form the transmit VAA (i.e., the D2D transmitter group) to multicast the information under the coordination of the BS. A CUE shares its downlink resource (e.g., subcarriers in OFDMA system) for the D2D link to reuse.  Also, in the system, a group of baleful UEs (i.e., eavesdroppers group) try to overhear the communication between D2D transceivers\footnote{For example, in commercial corporation or research institution, colleagues could share the internal information via VD2DM in a timely and convenient manner, meanwhile the VD2DM may be under the risk of eavesdrop attack from other UEs in the cellular system.}, thus the physical layer security is taken into consideration. Except for the BS, all nodes are equipped with one antenna, and the BS has $K$ antennas for its downlink transmission.
Let ${{\cal M}} \buildrel \Delta \over = {\rm{\{ 1,2,}} \ldots {\rm{,}}M{\rm{\} }}$ denote the set of D2D transmitters, ${{\cal N}} \buildrel \Delta \over = {\rm{\{ 1,2,}} \ldots {\rm{,}}N{\rm{\} }}$ denote the set of D2D receivers and ${{\cal L}} \buildrel \Delta \over = {\rm{\{ 1,2,}} \ldots {\rm{,}}L{\rm{\} }}$ denote the set of eavesdroppers. Define ${\bf{h}}_{D,n}^{}$ as the channel vector between D2D transmitters and D2D receiver $n$, ${\bf{h}}_l^{}$ as the channel vector between D2D transmitters and eavesdropper $l$, ${{\bf{h}}_C}$ as the channel vector between D2D transmitters and the CUE, ${{\bf{h}}_{B,n}}$ as the channel vector between BS and D2D receiver $n$, ${{\bf{h}}_{B,l}}$ as the channel vector between BS and eavesdropper $l$, and ${{\bf{h}}_{BC}}$ as the channel vector between BS and CUE, respectively. The Rayleigh block-fading channel model is assumed in this paper. Note that, the cellular network is considered as a kind of centralized network architecture and the eavesdroppers are also be considered as the registered active user in the cellular network, hence the BS can collect the channel
state information (CSI) of all links in the network\footnote{\color{blue}In LTE system, the BS and D2D transmitters could perform channel quality
indication estimation from the received sounding reference signal transmitted by
eavesdroppers then obtain $\bf{h}_C$ and $\bf{h}_l$, respectively. Further, D2D transmitters could report periodic or aperiodic report $\bf{h}_l$ to the BS
from the physical uplink control channel. Interested readers can read \cite{6231164} for the details.} \cite{8245811}. {\color{blue}Also, since the channels are quasi-static over each time block, the BS could precalculate
the the optimal beamforming vectors at the beginning of each time
block, then forwards the beamforming vector of the VAA to each D2D transmitters via physical
downlink control channel.}

We first consider the scenario in which each eavesdropper overhears the confidential information independently, and there has no extra information exchanges between eavesdroppers. The transmitted signal of the BS is

\begin{equation}
{{\bf{x}}_B} = {{\bf{w}}_B}{s_B} + {{\bf{n}}_B},
 \end{equation}
in which ${s_B} \sim {{\cal C}{\cal N}}\left( {0,1} \right)$ is the desired signal of CUE with beamforming vector ${{\bf{w}}_B}$, ${{\bf{n}}_B} \sim {{\cal C}{\cal N}}\left( {{\bf{0}},{\bf{Q}}} \right)$ is the AN vector to interfere eavesdroppers from overhearing with covariance matrix ${\bf{Q}} \in \mathbb{C} {^{K \times K}}$. The D2D transmitters form a VAA and employ the cooperative beamforming strategy to multicast the confidential information ${s_D} \sim {{\cal C}{\cal N}}\left( {0,1} \right)$. Specifically, the $m^{th}$ D2D transmitter multiplies the identical confidential information bearing signal $s_D$  with a complex weight ${w_m}$. Denote the cooperative beamforming vector by ${\bf{w}} \buildrel \Delta \over = {[{w_1},{w_2}, \ldots ,{w_M}]^T}$, then the received signals of D2D receiver $n$, eavesdropper $l$ and CUE can be respectively expressed as

\begin{equation}
y_n^{} = {\bf{h}}_{D,n}^H{\bf{w}}{s_D} + {\bf{h}}_{B,n}^H{\bf{w}}_B^{}{s_B}{\rm{ + }}{\bf{h}}_{B,n}^H{{\bf{n}}_B} + {n_n},n \in {{\cal N}},
 \end{equation}

\begin{equation}
{y_l} = {\bf{h}}_l^H{{\bf{w}}^{}}{s_D} + {\bf{h}}_{B,l}^H{\bf{w}}_B^{}{s_B} + {\bf{h}}_{B,l}^H{{\bf{n}}_B} + {n_l},l \in {{\cal L}},
 \end{equation}
and
\begin{equation}
y_{\rm{C}}^{} = {\bf{h}}_{BC}^H{\bf{w}}_B^{}{s_B} + {\bf{h}}_C^H{\bf{w}}{s_D}{\rm{ + }}{\bf{h}}_{BC}^H{{\bf{n}}_B} + {n_C},
 \end{equation}
where ${n_i} \sim {{\cal C}{\cal N}}\left( {0,1} \right),i \in \left\{ {{{\cal N}},{{\cal L}},C} \right\}$ is the Additive White Gaussian Noise (AWGN).

From (2), (3) and (4), the channel capacities of D2D receiver $n$, the eavesdropper $l$ to overhearing $s_D$, the eavesdropper $l$ to decoding $s_B$ and the CUE are respectively
\begin{equation}
{C_n}{\rm{ = }}\log \left( {1 + \frac{{{{\left| {{\bf{h}}_{D,n}^H{\bf{w}}} \right|}^2}}}{{{{\left| {{\bf{h}}_{B,n}^H{\bf{w}}_B^{}} \right|}^2} + {\rm{Tr}}\left( {{{\bf{H}}_{B,n}}{\bf{Q}}} \right) + 1}}} \right),
 \end{equation}

\begin{equation}
{C_l}{\rm{ = }}\log \left( {1 + \frac{{{{\left| {{\bf{h}}_l^H{\bf{w}}} \right|}^2}}}{{{{\left| {{\bf{h}}_{B,l}^H{\bf{w}}_B^{}} \right|}^2} +{\rm{Tr}}\left( {{{\bf{H}}_{B,l}}{\bf{Q}}} \right) + 1}}} \right),
 \end{equation}

 \begin{equation}
{C'_l}{\rm{ = }}\log \left( {1 + \frac{{{{\left| {{\bf{h}}_{B,l}^H{\bf{w}}_B^{}} \right|}^2}}}{{{{\left| {{\bf{h}}_{l}^H{\bf{w}}} \right|}^2}+{\rm{Tr}}\left( {{{\bf{H}}_{B,l}}{\bf{Q}}} \right)  + 1}}} \right),
 \end{equation}

 and
 \begin{equation}
{C_C}{\rm{ = }}\log \left( {1 + \frac{{{{\left| {{\bf{h}}_{BC}^H{\bf{w}}_B^{}} \right|}^2}}}{{{{\left| {{\bf{h}}_C^H{\bf{w}}} \right|}^2}+ {\rm{Tr}}\left( {{{\bf{H}}_{BC}}{\bf{Q}}} \right) + 1}}} \right),
 \end{equation}
in which where ${{\bf{H}}_{B,n}} \buildrel \Delta \over = {\bf{h}}_{B,n}^{}{\bf{h}}_{B,n}^H$, ${{\bf{H}}_{B,l}} \buildrel \Delta \over = {\bf{h}}_{B,l}^{}{\bf{h}}_{B,l}^H$ and ${{\bf{H}}_{BC}} \buildrel \Delta \over = {\bf{h}}_{BC}^{}{\bf{h}}_{BC}^H$

Let ${R_S}$ be the achievable secrecy rate of the D2D multicast communication and ${R_C}$ be the achievable rate of the CUE. Then an achievable rate region is given as the set of nonnegative rate pairs $\left( {{R_S},{R_C}} \right)$ satisfying

 \begin{equation}
{R_S} \le \mathop {\min }\limits_{n \in {{\cal N}}} {C_n} - \mathop {\max }\limits_{l \in{{\cal L}}} {C_l},
 \end{equation}
 \begin{equation}
{R_C} \le {C_C},
\end{equation}
and
\begin{equation}
{C'_l} < {R_C},
\end{equation}
in which $\mathop {\min }\limits_{n \in {{\cal N}}} {C_n}$ means the achievable rate of D2D multicast is determined by the receiver who has the minimum channel capacity. Also, $\mathop {\max }\limits_{l \in {{\cal L}}} {C_l}$ means the eavesdropping rate is determind by the eavesdropper who has the maximum channel capacity, since the confidential information cannot be overheard by any eavesdropper. ${C'_l} < {R_C}$ guarantees that the eavesdroppers could not perform non-orthogonal multiple access (NOMA) to decode and eliminate the $s_B$ first, hence the $s_B$ could jointly interfere with the eavesdroppers.

As can be seen in (5), (6), (7) and (8), the beamforming vector ${{\bf{w}}_B}$, cooperative beamforming vector ${\bf{w}}$  and the covariance matrix of AN, $\bf{Q}$ directly determine the achievable data rates, thus our work focuses on the design of ${{\bf{w}}_B}$, ${\bf{w}}$ and $\bf{Q}$, under an achievable rate region maximization (RRM) formulation with power constraint. The formulated RRM is a vector maximization problem, i.e. (12), in which $\chi$ is an arbitrarily number greater than zero.
In (12), $P_{max}$ and $P_B$ are the total power budgets of the VAA and the BS, respectively, the constraints C1 and C2 are the maximal power constraint of each D2D transmitter and the BS, C3, C4 and C5 are corresponding to (9), (10) and (11), respectively. The RRM problem in (12) is a non-convex vector optimization and thus difficult to solve. In the next section, we elaborate our approaches to tackle (12).
\begin{spacing}{1.00}
 \begin{equation}
 \begin{split}
({\rm{P1}})&\mathop {\max }\limits_{{{\bf{w}}_{\bf{B}}},{\bf{w}},{\bf{Q}}\succeq{\bf{0}},{R_S},{R_C}} \left( {{\rm{w}}{\rm{.r}}{\rm{.t}}{\rm{.\mathbb{R} }}_ + ^2} \right){\rm{ }}\left( {{\left[ {{R_S} ,0} \right]^ + },{\left[ {{R_C} ,0} \right]^ + }} \right)\\
 {\rm{s}}{\rm{.t.}}&{\rm{C1:   }}\left\| {{\bf{w}}[m]} \right\|_2^2 \le {P_{\max }}/M,m \in {{\cal M}}\\
&{\rm{C}}2:{\rm{ }}\left\| {{{\bf{w}}_B}} \right\|_2^2 \le {P_B}\\
&{\rm{  C3: }}\mathop {\min}\limits_{n \in {{\cal N}}} \log \left( {1 \!+ \!\frac{{{{\left| {{\bf{h}}_{D,n}^H{\bf{w}}} \right|}^2}}}{{{{\left| {{\bf{h}}_{B,n}^H{\bf{w}}_B^{}} \right|}^2}\!+\!{\rm{Tr}}\left( {{{\bf{H}}_{B,n}}{\bf{Q}}} \right) \!+ \!1}}} \right)\\
&{\;\;\;\;\;\;}- \mathop {\max}\limits_{l \in {{\cal L}}} \log \left( {1 \!+\! \frac{{{{\left| {{\bf{h}}_l^H{\bf{w}}} \right|}^2}}}{{{{\left| {{\bf{h}}_{B,l}^H{\bf{w}}_B^{}} \right|}^2}\!+ \!{\rm{Tr}}\left( {{{\bf{H}}_{B,l}}{\bf{Q}}} \right)\!+ \!1}}} \right) \ge {R_S}\\
 &{\rm{  C4: }}\log \left( {1 + \frac{{{{\left| {{\bf{h}}_{BC}^H{\bf{w}}_B^{}} \right|}^2}}}{{{{\left| {{\bf{h}}_C^H{\bf{w}}} \right|}^2} +{\rm{Tr}}\left( {{{\bf{H}}_{BC}}{\bf{Q}}} \right)+ 1}}} \right) - {R_C} \ge 0\\
 &{\rm{C5: }}{R_C} \ge \log \left( {1 + \frac{{{{\left| {{\bf{h}}_{B,l}^H{\bf{w}}_B^{}} \right|}^2}}}{{{{\left| {{\bf{h}}_l^H{\bf{w}}} \right|}^2} + {\rm{Tr}}\left( {{{\bf{H}}_{B,l}}{\bf{Q}}} \right) + 1}}} \right) + \chi ,l \in {{\cal L}}.
\end{split}
 \end{equation}
 \end{spacing}

\section{A Tractable Approach to Tackle the RRM Problem (WRMP)}

 A common approach for dealing with the vector optimization problem (12) is referred to as scalarization, whose basic idea is to transform the vector optimization into the weighted sum maximization optimization. In the rest of this section, we transform the RRM into a WRMP and then deal with it.

\subsection{Weighted Rate Maximization Problem}

  The target of WRMP is to maximize the total rate including the secrecy rate of D2D multicast communication and the rate of CUE under the total power constraints of both the D2D transmitters and the BS.

  From (5), (6), (7) and (8) we can see that, the transmitted signal vector from the D2D transmitter group, i.e., the VAA, ${\bf{w}}{s_D}$, not only could be overheard by eavesdroppers, but also interferes the CUE. Thus, the cooperative beamforming vector   ${\bf{w}}$ should be carefully designed for securing the D2D multicast communication while limiting the interference to the CUE. On the other hand, the transmitted signal vector from the BS, ${{\bf{w}}_B}{s_B}$, could be treated as a cooperative jamming signal to interfere eavesdroppers from overhearing the D2D multicast communication. Thus we need to design the beamforming vector ${{\bf{w}}_B}$ to increase the channel capacity of the CUE whereas decrease the channel capacities of eavesdroppers. Moreover the covariance matrix of the AN, $\bf{Q}$ also should be jointly designed to secure the communication while does not cause too much interference to other nodes.

  Since our aim is to maximize the total rate of the system, we then rewrite (12) into the following WRMP:
\begin{spacing}{1.00}
    \begin{equation}
  \begin{split}
{\rm{}}&\mathop {\max }\limits_{{\bf{w}},{{\bf{w}}_B},{\bf{Q}}\succeq {\bf{0}},{R_S},{R_C}} \alpha {R_S}{\rm{ + }}\left( {1 - \alpha } \right){R_C}\\
&{\rm{s}}{\rm{.t}}{\rm{. C1,C2,C3,C4,C5}}{\rm{.}}
\end{split}
    \end{equation}
\end{spacing}
  In (13), $\alpha  \in \left[ {0,1} \right]$ is the weighted coefficient, i.e., the larger $\alpha $ corresponds to that we tend to maximize the secrecy rate, vice versa.

  \subsection{Solving (13)}
  In the formulated WRMP, (13), although the constraints C1, C2 and the objective function are convex, the constraint C3 includes max-min and min-max structure simultaneously, which is non-convex and non-smoothness. To solve it, we next introduce a series of slack variables to transform it into convex expressions.

  First, we introduce two auxiliary variables $\varphi$ and $\beta$ to transform the originally constraint C3 into a smooth form. Specifically, let ${R_S} = {\left[ {\varphi  - \beta ,0} \right]^ + }$ \footnote{The ${\left[ { \bullet , \bullet } \right]^ + }$ operator ensures that ${R_S}$ is not less than zero, however for guaranteing that the objective function of (11) is linear, the ${\left[ { \bullet , \bullet } \right]^ + }$ operator is omitted and when $\varphi  - \beta < 0$, ${R_S}$ is set to zero.        }, (13) can be equivalently recast as
  \begin{spacing}{1.2}
      \begin{equation}
  \begin{split}
({\rm{P2}})&\mathop {\max }\limits_{{\bf{w}},{{\bf{w}}_B},{\bf{Q}}\succeq {\bf{0}},\varphi ,\beta ,{R_C}} {\rm{  }}\alpha \left( {\varphi  - \beta } \right) + \left( {1 - \alpha } \right){R_C}\\
{\rm{ s}}{\rm{.t.}}&{\rm{C1,C2,C4,C5}}\\
&{\rm{ C6: }}\log \left( {1 + \frac{{{{\left| {{\bf{h}}_{D,n}^H{\bf{w}}} \right|}^2}}}{{{{\left| {{\bf{h}}_{B,n}^H{\bf{w}}_B^{}} \right|}^2}+ {\rm{Tr}}\left( {{{\bf{H}}_{B,n}}{\bf{Q}}} \right)+ 1}}} \right) - \varphi  \ge 0,{\rm{ }}\forall n \in {{\cal N}}\\
&{\rm{         C7: }}\beta  - \log \left( {1 + \frac{{{{\left| {{\bf{h}}_l^H{\bf{w}}} \right|}^2}}}{{{{\left| {{\bf{h}}_{B,l}^H{\bf{w}}_B^{}} \right|}^2}+{\rm{Tr}}\left( {{{\bf{H}}_{B,l}}{\bf{Q}}} \right) + 1}}} \right) \ge 0,{\rm{ }}\forall l \in {{\cal L}}.
\end{split}
    \end{equation}
    \end{spacing}
 In (14), the original non-smooth constraint C3 is decomposed into groups of smooth constraints C6 and C7. However, constraints C6, C7, C4 and C5 are still non-convex and hard to deal with. Next, we respectively process C6, C7, C4 and C5 to transform them into convex constraints.

  \subsubsection{Convex approximation of C6}
  Let us introduce a group of auxiliary variable $\left\{ {a,b,c,d} \right\}$, then the constraint C6 can be further decomposed into the following constraints:
   \begin{spacing}{1.2}
   \begin{equation}
  \begin{array}{l}
{\rm{C}}8:{\rm{ log}}\left( {1 + a} \right) \ge \varphi  \Rightarrow 1 + a \ge {{\bf{e}}^\varphi }\\
{\rm{C}}9:{\rm{ }}\frac{{b_{}^2}}{{{c_{}}}} \ge a\\
{\rm{C10}}:{\rm{  }}{d_{}} \ge b_{}^2\\
{\rm{C11:   }}{\left| {{\bf{h}}_{D,n}^H{\bf{w}}} \right|^2} \ge d,\forall n \in {{\cal N}}\\
{\rm{C12: }}{c_{}} \ge {\left| {{\bf{h}}_{B,n}^H{\bf{w}}_B^{}} \right|^2} +{\rm{Tr}}\left( {{{\bf{H}}_{B,n}}{\bf{Q}}} \right) +1,\forall n \in {{\cal N}}.
\end{array}
     \end{equation}
      \end{spacing}
Now, C6 has been decomposed into a series of constraints. However, although C8, C10 and C12 are convex, they are not in the simplest form. Also, C9 and C11 are still non-convex. Let us first check C8, it is an exponential cone constraint and can be solved by existing nonlinear solvers (e.g., SDPT3 in CVX). To further reduce the computational complex, C8 can be approximated in terms of a series of second order conic (SOC) constraints. Let ${\bf{q}} \buildrel \Delta \over = {\left[ {{q_1},{q_2}, \ldots ,{q_{U + 4}}} \right]^T}$ be the introduced slack variables, C8 can be approximatively rewritten as
\begin{spacing}{1.2}
\begin{equation}
 {\rm{C8': }}\left\{ \begin{array}{l}
1 + a \ge {q_{n,U + 4}}\\
1 + {q_{U + 4}} \ge {\left\| {1 - {q_{U + 4}},2{q_{U + 3}}} \right\|_2}\\
1 + {q_i} \ge {\left\| {1 - {q_i},2{q_{i - 1}}} \right\|_2},i = 5, \ldots U + 3\\
{q_4} \ge 19/72 + {q_2} + {q_3}/24\\
1 + {q_3} \ge {\left\| {1 - {q_3},2{q_1}} \right\|_2}\\
1 + {q_2} \ge {\left\| {1 - {q_2},5/3 + \varphi /{2^U}} \right\|_2}\\
1 + {q_1} \ge {\left\| {1 - {q_1},2 + \varphi /{2^{U - 1}}} \right\|_2}
\end{array} \right.{\rm{ }}{\rm{,}}{ }
 \end{equation}\\
  \end{spacing}
\noindent{\color{blue}in which $U+4$ is the number of a series auxiliary variable $q$, and the approximate accuracy increases as $U+4$ increases.}

 It is easy to see that constraints C10 and C12 can be directly transformed to SOC forms as
 \begin{equation}
 {\rm{C10': }}\frac{{{d_{}} + 1}}{2} \ge {\left\| {{{\left[ {{b_{}},\frac{{d - 1}}{2}} \right]}^T}} \right\|_2}{\rm{ ,}}\forall n \in {{\cal N}},
  \end{equation}
and
 \begin{equation}
{\rm{C12': }}\frac{{{c_{}-{\rm{Tr}}\left( {{{\bf{H}}_{B,n}}{\bf{Q}}} \right)}}}{2} \ge {\left\| {{{\left[ {\left| {{\bf{h}}_{B,n}^H{\bf{w}}_B^{}} \right|,\frac{{c -{\rm{Tr}}\left( {{{\bf{H}}_{B,n}}{\bf{Q}}} \right)- 2}}{2}} \right]}^T}} \right\|_2},\forall n \in {{\cal N}},
  \end{equation}
  respectively. We next deal with C9. Based on the fact that the first-order Taylor expansion of a convex function is its lower bound, we employ the first-order Taylor series expansion of the left hand side of C9 around $b_{}^0$ and $c_{}^0$, i.e.,
   \begin{equation}
   {\rm{ }}\frac{{b_{}^2}}{c} \ge 2\left( {\frac{{b_{}^0}}{{c_{}^0}}} \right){b_{}} - {\left( {\frac{{b_{}^0}}{{c_{}^0}}} \right)^2}{c_{}}.
      \end{equation}

Then, C9 can be recast to
\begin{equation}
 {\rm{C}}9':2\left( {\frac{{b_{}^0}}{{c_{}^0}}} \right)b - {\left( {\frac{{b_{}^0}}{{c_{}^0}}} \right)^2}{c_{}} \ge a{}.
        \end{equation}
Note that, in ${\rm{C}}9'$, we use the lower bound of the left hand side of C8 to instead itself, thus the feasible region of ${\rm{C}}9'$ must be the subset of C9.

Let ${{\bf{H}}_{D,n}} \buildrel \Delta \over = {\bf{h}}_{D,n}^{}{\bf{h}}_{D,n}^H$, the left hand side of C11 can be written to ${{\bf{w}}^H}{{\bf{H}}_{D,n}}{\bf{w}}$, which is a semipositive definite quadratic form in complex domain. Following \cite{complexvector}, we define $g\left( {{\bf{w}},{{\bf{w}}^H}} \right) = {{\bf{w}}^H}{{\bf{H}}_{D,n}}{\bf{w}}$, according to eq.(15) in \cite{complexvector}, giving a differential of ${\bf{w}}$, $\delta {\bf{w}}$, we have
\begin{equation}
\delta g{\rm{ = }}{\left( {{\nabla _{\bf{w}}}g} \right)^T}\delta {\bf{w}} + {\left( {{\nabla _{{{\bf{w}}^*}}}g} \right)^T}\delta {{\bf{w}}^*} = 2{\mathop{\rm Re}\nolimits} \left( {{{\left( {{\nabla _{{{\bf{w}}^*}}}g} \right)}^H}\delta {\bf{w}}} \right),
\end{equation}
in which ${{\nabla _{\bf{w}}}g}$ and ${{\nabla _{{{\bf{w}}^*}}}g}$ are the generalized complex gradient and the conjugate complex gradient of $g\left( {{\bf{w}},{{\bf{w}}^H}} \right)$ defined in \cite{complexvector}, respectively.

With the help of two types of gradients, it is easy to calculate that
\begin{equation}
{\nabla _{{{\bf{w}}^*}}}g = {{\bf{H}}_{D,n}}{\bf{w}}.
\end{equation}

Hence the first-order Taylor series expansion of the left hand side of C11 around ${\bf{w}}_{}^0$ is
\begin{equation}
{\bf{w}}_{}^{0H}{{\bf{H}}_{D,n}}{{\bf{w}}^0}{\rm{ + }}2{\mathop{\rm Re}\nolimits} \left( {{\bf{w}}_{}^{0H}{{\bf{H}}_{D,n}}\left( {{\bf{w}} - {{\bf{w}}^0}} \right)} \right){\rm{ = }}2{\mathop{\rm Re}\nolimits} \left( {{\bf{w}}_{}^{0H}{{\bf{H}}_{D,n}}{\bf{w}}} \right) - {\bf{w}}_{}^{0H}{{\bf{H}}_{D,n}}{{\bf{w}}^0},
\end{equation}
and then C11 can be approximated as

\begin{equation}
{\rm{C}}11':2{\mathop{\rm Re}\nolimits} \left( {{\bf{w}}_{}^{0H}{{\bf{H}}_{D,n}}{\bf{w}}} \right) - {\bf{w}}_{}^{0H}{{\bf{H}}_{D,n}}{{\bf{w}}^0} \ge d,\forall n \in {{\cal N}}.
\end{equation}
Thus far, the non-convex constraint C6 has been decomposed to a series of convex constraints.

\subsubsection{Convex approximation of C7}

 By introducing another group of auxiliary variable $\left\{ {e,f} \right\}$, C7 can be decomposed into the following constraints:
  \begin{spacing}{1.2}
\begin{equation}
\begin{array}{l}
{\rm{C}}13:{\rm{ }}\beta  - {\rm{log}}\left( {1 + e} \right) \ge 0\\
{\rm{C}}14:{\rm{ }}e \ge \frac{{{{\left| {{\bf{h}}_l^H{\bf{w}}} \right|}^2}}}{f},{\rm{ }}\forall l \in {{\cal L}}\\
{\rm{C}}15:{\rm{ }}{\left| {{\bf{h}}_{B,l}^H{\bf{w}}_B^{}} \right|^2} +{\rm{Tr}}\left( {{{\bf{H}}_{B,l}}{\bf{Q}}} \right)+ 1 \ge {f_{}},{\rm{ }}\forall l \in {{\cal L}}.
\end{array}
\end{equation}
 \end{spacing}
In (25), C13 and C15 are still non-convex and C14 can be rewritten to SOC form directly:
\begin{equation}
{\rm{C}}14':{\rm{ }}\frac{{{e_{}}{\rm{ + }}{f_{}}}}{2} \ge {\left\| {{{\left[ {\frac{{e - {f_{}}}}{2},\left| {{\bf{h}}_l^H{\bf{w}}} \right|} \right]}^T}} \right\|_2},{\rm{ }}\forall l \in{{\cal L}}.
\end{equation}
By preforming the first-order Taylor series expansion, C13 and C15 can be approximated as
\begin{equation}
{\rm{C}}13':{\rm{ }}\beta  - {\rm{log}}\left( {1 + e_{}^0} \right) - \frac{1}{{1 + e_{}^0}}\left( {e_{}^{} - e_{}^0} \right) \ge 0,
\end{equation}
and
\begin{equation}
{\rm{C}}15':2{\mathop{\rm Re}\nolimits} \left( {{\bf{w}}_B^{0H}{{\bf{H}}_{B,l}}{\bf{w}}_B^{}} \right) - {\bf{w}}_B^{0H}{{\bf{H}}_{B,l}}{\bf{w}}_B^0{\rm{ + }}{\rm{Tr}}\left( {{{\bf{H}}_{B,l}}{\bf{Q}}} \right)+1 \ge {f_{}},{\rm{ }}\forall l \in {{\cal L}},
\end{equation}
respectively. With the series of constrains ${\rm{C}}13'$, ${\rm{C}}14'$and ${\rm{C}}15'$, C7 has been approximated to a convex constraint.

\subsubsection{Convex approximation of C4 and C5}
Note that C4 and C6 have the similar formation, thus with the same approach of C6 above, C4 could be approximated to the following constraints C16, C17, C18, C19 and C20 corresponding to ${\rm{C}}8'$,${\rm{C}}9'$,${\rm{C}}10'$,${\rm{C}}11'$ and ${\rm{C}}13'$, respectively.

Similarly, C5 could be approximated to C21, C22 and C23 with the same approach of C7 above. For briefly, we have omitted the details of constraints C16$ {\rm{\sim}}$C23.

Now, all non-convex constraints have been approximated to convex constraints, the approximated version of WRMP is given by
\begin{equation}
\begin{array}{l}
({\rm{P3}})\mathop {\max }\limits_{\scriptstyle{\bf{w}},{{\bf{w}}_B},{\bf{Q}}\succeq {\bf{0}},{R_S},{R_C},\varphi ,\beta ,a,{b},\hfill\atop
\scriptstyle{c},{d},e,{f},{\bf{q}}\hfill} \alpha \left( {\varphi  - \beta } \right){\rm{ + }}\left( {1 - \alpha } \right){R_C}\\
\;\;\;\;\;\;\;\;\;\;\;\;{\rm{s}}{\rm{.t}}{\rm{. C1,C2,C8',C9',C10',C11',C12',}}\\
{\rm{ C13',C14',C15',C16,C17,C18,C19,C20,C21,C22,C23}}{\rm{.}}
\end{array}
\end{equation}
Note that, for briefly, the auxiliary variables for approximating C4 and C5 are omitted in ${\rm{P3}}$. It easy to see that ${\rm{P3}}$ is a convex problem, which can be efficiently solved by using interior point methods, or standard convex optimization tool, e.g., CVX. The detailed procedure is summarized in Algorithm 1.

\begin{algorithm}[t]
\caption{Solving the WRMP in (13).}

\label{alg1}
\begin{algorithmic}[1]
\STATE \textbf{Initialization:}

$\bullet$ Setting the initial values $b_{}^0$, $c_{}^0$, $e_{}^0$, ${\bf{w}}_{}^0$ and ${\bf{w}}_B^0$.

$\bullet$ Setting maximum number of iterations $J_{}^{\max }$, iteration counter $j=1$ and iteration termination accuracy $\psi  > 0$.
\STATE \textbf{Repeat}
 \STATE Solving the ${\rm{WRMP'}}$ with current $b_{}^0$, $c_{}^0$, $e_{}^0$, ${\bf{w}}_{}^0$, ${\bf{w}}_B^0$ and obtaining the optimal $b_{}^*$, $c_{}^*$, $e_{}^*$, ${\bf{w}}_{}^*$, ${\bf{w}}_B^*$.
 \IF {${\left\| {{\bf{w}}_{}^{\rm{*}} - {\bf{w}}_{}^0} \right\|_2} + {\left\| {{\bf{w}}_B^{\rm{*}} - {\bf{w}}_B^0} \right\|_2} \le \psi $.}
      \STATE The optimal solutions of ${\rm{WRMP'}}$ are ${\bf{w}}_{}^*$ and ${\bf{w}}_B^*$.
      \STATE \textbf{Break}
 \ELSE
      \STATE Setting $b_{}^0 = b_{}^{\rm{*}}$, $c_{}^0 = c_{}^{\rm{*}}$, $e _{}^0 = e_{}^{\rm{*}}$, ${\bf{w}}_{}^0 = {\bf{w}}_{}^*$ and ${\bf{w}}_{B}^0 = {\bf{w}}_{B}^*$.
 \ENDIF
 \STATE $j=j+1$
 \STATE \textbf{Until} $j>J_{}^{\max }$
\end{algorithmic}
\end{algorithm}

{\it Complexity and convergence:} The complexity of interior point method for solving a convex optimization problem (i.e., line 3 in Algorithm 1) is $O\left( {{\sigma ^{0.5}}(\sigma  + \varsigma ){\varsigma ^2}} \right)$\cite{8663343}, in which $\sigma {\rm{ = }}{2U + N + 4L + 19}$ is the number of inequality constraints and $\varsigma=2K + 4M + 2U $$+ 24$ is the number of optimization variables. Combining with the fact that the iterative number of Algorithm 1 is bounded by $J_{}^{\max }$, we have the total computational complexity of Algorithm 1 is $O\left( J_{}^{\max }{{\sigma ^{0.5}}(\sigma  + \varsigma ){\varsigma ^2}} \right)$.

 The convergence of Algorithm 1 could refer to Proposition 2 in \cite{8245811}.

\begin{remark}
{\color{blue}Our formulated WRMP in both the normal case and the worst case does not include the quality of service (QoS) ensuring constraints which limit the achievable rate of CUE and the achievable secrecy rate of D2DM at least to a certain value, that is because the QoS constraints may lead to the WRMP being unsolvable. In practice, if the achievable secrecy rate of the solution, $R_S$ is too small, we could increase $\alpha $ appropriately, vice visa.}
\end{remark}

\section{Extension to the Worst Case}

In Section III above, we only considered the scenario in which multiple eavesdroppers overhear the confidential information independently. However, in the worst case that the eavesdroppers may collude and form another VAA to jointly process the eavesdropped information (such as maximum ratio combination and MMSE receiver) to increase the overhearing capacity. We next make efforts to investigate the PLS scheme in such a scenario.

Since the eavesdroppers are ganged up together and form an eavesdrop VAA, the group of eavesdroppers i.e., the eavesdrop VAA could be treated as a virtual multi-antenna eavesdropper which is equipped with $L$ antennas. In the rest of this paper, unless particularly stated, we use Eve to represent the so-called eavesdrop VAA, i.e., the equivalent $L$-antenna eavesdropper.

Since the eavesdroppers form the eavesdrop VAA, the channel vector from D2D transmitter group to the $l^{th}$ eavesdropper, ${\bf{h}}_l^{}$ could be treated as the channel vector from D2D transmitter group to the $l^{th}$ antenna of Eve. We use ${{\bf{G}}_E} \buildrel \Delta \over = \left[ {{\bf{h}}_1^{},{\bf{h}}_2^{}, \ldots ,{\bf{h}}_L^{}} \right]$ represent the channel matrix between D2D transmitter group and Eve. Similarly, we use ${{\bf{G}}_B} \buildrel \Delta \over = \left[ {{{\bf{h}}_{B,1}}{{\bf{h}}_{B,2}},...,{{\bf{h}}_{B,L}}} \right]$ represent the channel matrix between BS and Eve. Then the received signals of D2D receiver $n$, Eve and CUE can be expressed as
 \begin{equation}
y_n^{} = {\bf{h}}_{D,n}^H{\bf{w}}{s_D} + {\bf{h}}_{B,n}^H{\bf{w}}_B^{}{s_B}{\rm{ + }}{\bf{h}}_{B,n}^H{{\bf{n}}_B} + {n_n},n \in {{\cal N}},
 \end{equation}
 \begin{equation}
{{\bf{y}}_E} = {\bf{G}}_E^H{{\bf{w}}^{}}{s_D} + {\bf{G}}_B^H{\bf{w}}_B^{}{s_B} + {\bf{G}}_B^H{{\bf{n}}_B} + {{\bf{n}}_E},
 \end{equation}
 and
  \begin{equation}
 y_{\rm{C}}^{} = {\bf{h}}_{BC}^H{\bf{w}}_B^{}{s_B}{\rm{ + }}{\bf{h}}_{BC}^H{{\bf{n}}_{\rm{B}}} + {\bf{h}}_C^H{\bf{w}}{s_D} + {n_C},
  \end{equation}
respectively, where ${{\bf{n}}_E} \sim CN\left( {{\bf{0}},{{\bf{I}}_L}} \right)$ is the AWGN vector at Eve. Similar with the normal case, from (30), (31) and (32), the channel capacities of D2D receiver $n$, the eavesdropper Eve to overhearing $s_D$\footnote{\color{blue} The channel capacity of Eve is obtained by calculating the mean mutual information which is the upper bound of the achievable rate of Eve, regardless of which receiving filter Eve used. Hence, the achievable secrecy rate of the worst case is a lower bound.}, the eavesdropper $l$ to decoding $s_B$ and the CUE are
  \begin{equation}
 C_n^{worst}{\rm{ = }}\log \left( {1 + \frac{{{{\bf{w}}^H}{\bf{H}}_{D,n}^{}{\bf{w}}}}{{{\bf{w}}_B^H{\bf{H}}_{B,n}^{}{\bf{w}}_B^{} + {\bf{h}}_{B,n}^H{\bf{Qh}}_{B,n}^{} + 1}}} \right),
   \end{equation}
  \begin{equation}
C_E^{worst}{\rm{\!=\!}}\log \left(\!{\det \left(\! {{{\bf{I}}_L}\! +\! \frac{{{\bf{G}}_E^H{\bf{w}}{{\bf{w}}^H}{\bf{G}}_E^{}}}{{{\bf{G}}_B^H{\bf{w}}_B^{}{\bf{w}}_B^H{\bf{G}}_B^{}\! + \! {\bf{G}}_B^H{\bf{QG}}_B^{}\! + \!{{\bf{I}}_L}}}} \!\right)}\!\right),
   \end{equation}
    \begin{equation}
{C{_E^{worst}}}^{\prime} {\rm{\!=\!}}\log \left( {\det \left( {{{\bf{I}}_L} + \frac{{{\bf{G}}_B^H{\bf{w}}_B^{}{\bf{w}}_B^H{\bf{G}}_B^{}}}{{{\bf{G}}_E^H{\bf{w}}{{\bf{w}}^H}{\bf{G}}_E^{} + {\bf{G}}_B^H{\bf{QG}}_B^{} + {{\bf{I}}_L}}}} \right)} \right),
   \end{equation}
  and
   \begin{equation}
C_C^{worst}{\rm{ = }}\log \left( {1 + \frac{{{\bf{w}}_B^H{\bf{H}}_{BC}^{}{\bf{w}}_B^{}}}{{{\bf{h}}_{BC}^H{\bf{Qh}}_{BC}^{} + {\bf{w}}_{}^H{\bf{H}}_C^{}{\bf{w}}_{}^{} + 1}}} \right),
   \end{equation}
respectively, where ${\bf{H}}_C^{} \buildrel \Delta \over = {\bf{h}}_C^{}{\bf{h}}_C^H$. It is important to note that since the eavesdroppers form another VAA, the eavesdrop channel has transformed from MISO to MIMO, and the corresponding channel capacity becomes (34) and (35).

The achievable rate region is then defined as the set of nonnegative rate pairs $\left( {R_S^{worst},R_C^{worst}} \right)$ satisfying
 \begin{equation}
R_S^{worst} \le \mathop {\min }\limits_{n \in {{\cal N}}} C_n^{worst} - C_E^{worst},
\end{equation}
 \begin{equation}
  R_C^{worst} \le C_C^{worst},
  \end{equation}
  and
  \begin{equation}
{C{_E^{worst}}}^{\prime} < {\rm{ }}R_C^{worst}.
\end{equation}

 {\color{blue}In the worst case, to reduce computational complexity, we discuss the WRMP in two subcases and formulate corresponding optimization problems. Note that, different from the WRMP in the normal case which include constraints C5 to guarantee that the eavesdroppers could not perform NOMA to decode and eliminate the $s_B$ first. In both subcases, we do not consider (39) into the optimization problems, the detailed reasons please see Remark 3 at last of this section.

In subcase 1, we consider that Eve could not decode and eliminate the $s_B$ first and formulate the corresponding sub-WRMP, ${\rm{P4\_1}}$. On the contrary, in subcase 2, we consider that Eve could decode and eliminate the $s_B$ first and formulate the corresponding sub-WRMP, ${\rm{P4\_2}}$.}

We first consider subcase 1 and formulate the following ${\rm{P4\_1}}$ to maximize the weighted sum of $R_S^{worst}$ and $R_C^{worst}$

\begin{equation}
 \begin{split}
({\rm{P4\_1}})&\mathop {\max }\limits_{\scriptstyle{\bf{w}},{{\bf{w}}_B},{\bf{Q}}\succeq {\bf{0}},\varphi _{}^{worst}\hfill\atop
\scriptstyle{\beta ^{worst}},R_C^{worst}\hfill} {\rm{  }}\alpha \left( {\varphi _{}^{worst} - {\beta ^{worst}}} \right) + \left( {1 - \alpha } \right)R_C^{worst}\\
{\rm{s}}{\rm{.t}}&{\rm{.C1,C2}}\\
&{\rm{C}}4':\log \left( {1 + \frac{{{\bf{w}}_B^H{\bf{H}}_{BC}^{}{\bf{w}}_B^{}}}{{{\bf{h}}_{BC}^H{\bf{Q}}_{}^{}{\bf{h}}_{BC}^{} + {\bf{w}}_{}^H{\bf{H}}_C^{}{\bf{w}}_{}^{} + 1}}} \right) \ge R_C^{worst}\\
&{\rm{C6': }}\log \left( {1 + \frac{{{{\bf{w}}^H}{\bf{H}}_{D,n}^{}{\bf{w}}}}{{{\bf{w}}_B^H{\bf{H}}_{B,n}^{}{\bf{w}}_B^{} + {\bf{h}}_{B,n}^H{\bf{Qh}}_{B,n}^{} + 1}}} \right)- \varphi _{}^{worst} \ge 0,{\rm{ }}\forall n \in {{\cal N}}\\
&{\rm{ C7':}}{\beta ^{worst}}\!-\log\! \left(\! {\det\! \left(\! {{{\bf{I}}_L}\! +\! \frac{{{\bf{G}}_E^H{\bf{w}}{{\bf{w}}^H}{\bf{G}}_E^{}}}{{{\bf{G}}_B^H{\bf{w}}_B^{}{\bf{w}}_B^H{\bf{G}}_B^{}\! +\! {\bf{G}}_B^H{\bf{QG}}_B^{} \!+ \!{{\bf{I}}_L}}}}\! \right)} \!\right) \ge 0.
\end{split}
 \end{equation}
where $R_S^{worst} = \varphi _{}^{worst} - {\beta ^{worst}}$ and $\alpha$ is the weighted coefficient. ${\rm{C}}4'$, ${\rm{C}}6'$ and ${\rm{C}}7'$ are evolved from C4, C5 and C7 in (14) respectively, because of the different expressions of the channel capacity of CUE, D2D receivers and Eve.

{\color{blue}Note that, ${\rm{P4\_2}}$ is basically the same as ${\rm{P4\_1}}$, by replacing ${\rm{ C7'}}$ with
  \begin{equation}
{\rm{ C7'\_2:}}{\beta ^{worst}}\!-\log\! \left(\! {\det\! \left(\! {{{\bf{I}}_L}\! +\! \frac{{{\bf{G}}_E^H{\bf{w}}{{\bf{w}}^H}{\bf{G}}_E^{}}}{{ {\bf{G}}_B^H{\bf{QG}}_B^{} \!+ \!{{\bf{I}}_L}}}}\! \right)} \!\right) \ge 0.
\end{equation}
${\rm{P4\_2}}$ can be obtained. In ${\rm{ C7'\_2:}}$, only the AN could interfere with Eve since Eve could decode and eliminate
$s_B$ first.}

We first make effort to solve ${\rm{P4\_1}}$, it is easy to see that ${\rm{C}}4'$ and ${\rm{C}}6'$ have the same structure as C4 and C6, respectively, and the technique used to transform ${\rm{C}}4$ and ${\rm{C}}6$ to series of convex constraints in Section III.B could still be used to deal with ${\rm{C}}4'$ and ${\rm{C}}6'$. However, due to the determinant operation, ${\rm{C}}7'$ in (40) is a non-convex constraint and non-trivial to deal with.
 \subsection{A Semidefinite Relaxation Based Approach for handling ${\rm{C}}7'$ }

Observing that, ${{\bf{x}}^H}{\bf{Rx}} = {\rm{Tr}}\left( {{\bf{Rx}}{{\bf{x}}^H}} \right)$, when ${\bf{x}}$ is a complex vector and ${\bf{R}}$ is Hermitian, to handle problem (40), we define ${{\bf{W}}_B} \buildrel \Delta \over = {\bf{w}}_B^{}{\bf{w}}_B^H$, ${\bf{W}} \buildrel \Delta \over = {\bf{ww}}_{}^H$ and ignore the non-convex constraints ${\rm{Rank}}\left( {\bf{W}} \right) \le 1$, ${\rm{Rank}}\left( {{{\bf{W}}_B}} \right) \le 1$ \footnote{When the optimal solution $\left( {{\bf{W}^ * },{\bf{W}}_B^ * } \right)$ to (45) satisfies ${\rm{Rank}}\left( {\bf{W}} \right) \le 1$, ${\rm{Rank}}\left( {{{\bf{W}}_B}} \right) \le 1$, the corresponding maximum weight sum rate could be obtained via single-stream transmit beamforming, which facilitates the implementation of physically realizable transceiver with low complexity. However the Rank constrains is non-convex and hard to deal with, here we omit them and will discuss it in Section IV-D.} to obtain a relaxed problem version of (40) as follow
\begin{spacing}{1.00}
\begin{equation}
 \begin{split}
({\rm{P5}})&\mathop {\max }\limits_{\scriptstyle{\bf{W}}\succeq {\bf{0}},{{\bf{W}}_B} \succeq {\bf{0}},{\bf{Q}}\succeq {\bf{0}},\varphi _{}^{worst}\hfill\atop
\scriptstyle{\beta ^{worst}},R_C^{worst}\hfill} {\rm{  }}\alpha \left( {\varphi _{}^{worst} - {\beta ^{worst}}} \right) + \left( {1 - \alpha } \right)R_C^{worst}\\
&{\rm{ C1': }}{\bf{W}}[m] \le {P_{\max }}/M,\forall m \in {{\cal M}}\\
&{\rm{C2': Tr}}\left( {{{\bf{W}}_B}} \right) + {\mathop{\rm Tr}\nolimits} \left( {\bf{Q}} \right) \le {P_B}\\
&{\rm{C}}4'':\log \!\left(\! {1\! +\! \frac{{{\mathop{\rm Tr}\nolimits}\! \left( \!{{\bf{H}}_{BC}^{}{{\bf{W}}_B}} \!\right)}}{{{\mathop{\rm Tr}\nolimits}\! \left( \!{{\bf{Qh}}_{BC}^{}{\bf{h}}_{BC}^H} \!\right)\! +\! {\mathop{\rm Tr}\nolimits} \!\left( \!{{\bf{H}}_C^{}{\bf{W}}} \!\right)\! +\! 1}}}\! \right)\! \ge\! R_C^{worst}\\
&{\rm{C6'': }}\log \left( {1 + \frac{{{\mathop{\rm Tr}\nolimits} \left( {{\bf{H}}_{D,n}^{}{\bf{W}}} \right)}}{{{\mathop{\rm Tr}\nolimits} \left( {{\bf{H}}_{B,n}^{}{{\bf{W}}_B}} \right) + {\mathop{\rm Tr}\nolimits} \left( {{\bf{QH}}_{B,n}^{}} \right) + 1}}} \right)- \varphi _{}^{worst} \ge 0,{\rm{ }}\forall n \in {{\cal N}}\\
&{\rm{ C7'':}}{\beta ^{worst}}\!- \log\!\left(\! {\det \left( \!{{{\bf{I}}_L}\! +\! \frac{{{\bf{G}}_E^H{\bf{WG}}_E^{}}}{{{\bf{G}}_B^H{{\bf{W}}_B}{\bf{G}}_B^{}\! + \!{\bf{G}}_B^H{\bf{QG}}_B^{} \!+ \!{{\bf{I}}_L}}}} \!\right)}\! \right)\! \ge \!0.
\end{split}
 \end{equation}
 \end{spacing}
In (46), ${\rm{C7''}}$ is still hard to handle. Before we deal with it, we first introduce the following lemma whose proof can be found in \cite{6482662}:
\begin{lemma}
\begin{equation}
\begin{array}{l}
{\rm{C}}7'' \Rightarrow {\rm{  C24: }}\left(\! {{{\bf{e}}^{{\beta ^{worst}}}}\! - 1}\! \right)\left(\! {{\bf{G}}_B^H{{\bf{W}}_B}{\bf{G}}_B^{} \!+ \!{\bf{G}}_B^H{\bf{QG}}_B^{}\! + \!{{\bf{I}}_L}}\! \right) - {\bf{G}}_E^H{\bf{WG}}_E^{}\succeq {\bf{0}}.
\end{array}
 \end{equation}
For any ${\bf{G}}_E^{} \in _{}^{M \times L}$, ${\bf{G}}_B^{} \in _{}^{K \times L}$, ${\bf{W}}_B^{} \succeq {\bf{0}}$, ${\bf{W}}^{} \succeq {\bf{0}}$ and ${\bf{Q}} \succeq  {\bf{0}}$. Furthermore, the equivalence in (47) holds if ${\mathop{\rm Rank}\nolimits} \left( {\bf{W}} \right) \le 1$.
\end{lemma}

Replacing ${\rm{C7''}}$ with C24, (42) becomes:
\begin{equation}
\begin{array}{l}
({\rm{P6}})\mathop {\max }\limits_{\scriptstyle{\bf{W}}\succeq {\bf{0}},{{\bf{W}}_B} \succeq {\bf{0}},{\bf{Q}}\succeq {\bf{0}},\hfill\atop
\scriptstyle\varphi _{}^{worst},{\beta ^{worst}},R_C^{worst}\hfill} {\rm{  }}\alpha \left( {\varphi _{}^{worst} - {\beta ^{worst}}} \right) + \left( {1 - \alpha } \right)R_C^{worst}\\
\;\;\;\;\;\;\;\;\;\;\;\;\;\;\;\;\;\;\;\;\;\;\;{\rm{s}}{\rm{.t}}{\rm{. C1', C2', C}}4'',{\rm{C6'',C24}}{\rm{.}}
\end{array}
 \end{equation}
 It is obvious that when ${\beta ^{worst}}$ is fixed, C24 becomes a linear matrix inequality which is a convex constraint. Hence (44) could be deposed into a two-level optimization: the inner-level optimization is convex with a fixed ${\beta ^{worst}}$, and the
outer-level is a single-variable optimization with respect to ${\beta ^{worst}}$.

 \subsection{The Inner-Level Optimization }
 With a fixed ${\beta ^{worst}}$, the inner-level optimization can be written as

 \begin{equation}
\begin{array}{l}
({\rm{P7}})\mathop {\max }\limits_{\scriptstyle{\bf{W}}\succeq {\bf{0}},{{\bf{W}}_B} \succeq {\bf{0}},{\bf{Q}}\succeq {\bf{0}},\hfill\atop
\scriptstyle\varphi _{}^{worst},R_C^{worst}\hfill} {\rm{  }}\alpha \left( {\varphi _{}^{worst} - {\beta ^{worst}}} \right) + \left( {1 - \alpha } \right)R_C^{worst}\\
\;\;\;\;\;\;\;\;\;\;\;\;\;\;\;\;\;\;\;\;\;\;\;{\rm{s}}{\rm{.t}}{\rm{. C1', C2', C}}4'',{\rm{C6'',C24}}{\rm{.}}
\end{array}
 \end{equation}
 In (45), although ${\rm{C1'}}$, ${\rm{C2'}}$ and C24 are convex constraint, we still need to deal with the non-convex constraints   ${\rm{C}}4''$ and ${\rm{C}}6''$. Note that, the convex approximation approach in Section III.B could be used to transfer ${\rm{C}}4''$ and ${\rm{C}}6''$ into convex form. However, observing that the expressions inside the logarithm of ${\rm{C}}4''$ and ${\rm{C}}6''$ are in the form of linear fraction, hence, in the following, we give an easier approach that does not need complex domain Taylor expansion.
  \subsubsection{ Convex approximation of ${\rm{C}}4''$  }
  By introducing auxiliary variable $r_{}^{worst}$, ${\rm{C}}4''$can be decomposed into
  \begin{equation}
 {\rm{C}}25:\log \left( {1 + r_{}^{worst}} \right) \ge R_C^{worst} \Leftrightarrow 1 + r_{}^{worst} \ge {{\bf{e}}^{R_C^{worst}}},
  \end{equation}
 and
 \begin{equation}
 \begin{array}{l}
{\rm{C}}26:\frac{{{\mathop{\rm Tr}\nolimits} \left( {{\bf{H}}_{BC}^{}{{\bf{W}}_B}} \right)}}{{{\mathop{\rm Tr}\nolimits} \left( {{\bf{QH}}_{BC}^{}} \right) + {\mathop{\rm Tr}\nolimits} \left( {{\bf{H}}_C^{}{\bf{W}}} \right) + 1}} \ge r_{}^{worst} \Leftrightarrow {\rm{}}{\mathop{\rm Tr}\nolimits} \!\left(\! {{\bf{H}}_{BC}^{}{{\bf{W}}_B}}\! \right) \!\ge\! r_{}^{worst}\left(\! {{\mathop{\rm Tr}\nolimits} \left( \!{{\bf{QH}}_{BC}^{}}\! \right)\! +\! {\mathop{\rm Tr}\nolimits} \left(\! {{\bf{H}}_C^{}{\bf{W}}} \!\right) \!+ \!1} \!\right).
\end{array}
  \end{equation}
 Like C8 in (15), C21 is an exponential cone constraint and can be approximated in terms of a series of SOC constraints with the same formation of ${\rm{C}}8'$. For brevity, we denote the series of SOC constraints derived from C25 by ${\rm{C}}25'$ and omitting its detailed expression.

 Then, arithmetic geometry mean (AGM) inequality is applied to get an approximation of the non-convex constraint, i.e., C26. We first introduce the AGM inequality. The AGM inequality is described as: $xy \le 0.5\left( {\eta x} \right)_{}^2 + 0.5\left( {y/\eta } \right)_{}^2$, where $\eta  \ne 0$ and the equality holds if and only if when $\eta  = \sqrt {y/x} $. Based the AGM inequality, C26 can be approximately recast as
 \begin{equation}
\begin{array}{l}
{\rm{C26'}}:{\mathop{\rm Tr}\nolimits} \left( {{\bf{H}}_{BC}^{}{{\bf{W}}_B}} \right) \ge \frac{1}{2}{\left( {{\eta _0}r_{}^{worst}} \right)^2}+\frac{1}{2}{\left( {\frac{{{\mathop{\rm Tr}\nolimits} \left( {{\bf{QH}}_{BC}^{}} \right) + {\mathop{\rm Tr}\nolimits} \left( {{\bf{H}}_C^{}{\bf{W}}} \right) + 1}}{{{\eta _0}}}} \right)^2},
\end{array}
  \end{equation}
  which is convex.

  \subsubsection{ Convex approximation of ${\rm{C}}6''$  }

  With the auxiliary variable $a_{}^{worst}$, we decompose ${\rm{C6''}}$ into
  \begin{equation}
  {\rm{C}}27:\log \left( {1 + a_{}^{worst}} \right) \ge \varphi _{}^{worst} \Leftrightarrow 1 + a_{}^{worst} \ge {{\bf{e}}^{\varphi _{}^{worst}}},
    \end{equation}
  and
   \begin{equation}
\begin{array}{l}
{\rm{C}}28:\frac{{{\rm{Tr}}\left( {{\bf{H}}_{D,n}^{}{\bf{W}}} \right)}}{{{\rm{Tr}}\left( {{\bf{H}}_{B,n}^{}{{\bf{W}}_B}} \right) + {\rm{Tr}}\left( {{\bf{QH}}_{B,n}^{}} \right) + 1}} \ge a_{}^{worst} \Leftrightarrow \\
{\rm{       Tr}}\left( {{\bf{H}}_{D,n}^{}{\bf{W}}} \right) \ge a_{}^{worst}\left( {{\rm{Tr}}\left( {{\bf{H}}_{B,n}^{}{{\bf{W}}_B}} \right) + {\rm{Tr}}\left( {{\bf{QH}}_{B,n}^{}} \right) + 1} \right),\forall n \in N.
\end{array}
   \end{equation}

Similar with how we process C25, C27 is replaced by a series of SOC constraints, i.e., ${\rm{C27'}}$ and we omit the detialed expression of it.
With the help of AGM inequality, C28 can also be approximated to a convex constraint:
\begin{equation}
  \begin{array}{l}
{\rm{C}}28':{\mathop{\rm Tr}\nolimits} \left( {{\bf{H}}_{D,n}^{}{\bf{W}}} \right) \ge \frac{1}{2}{\left( {{\eta _n}a_{}^{worst}} \right)^2}+\frac{1}{2}{\left( {\frac{{{\mathop{\rm Tr}\nolimits} \left( {{\bf{H}}_{B,n}^{}{{\bf{W}}_B}} \right) + {\mathop{\rm Tr}\nolimits} \left( {{\bf{QH}}_{B,n}^{}} \right) + 1}}{{{\eta _n}}}} \right)^2},{\rm{ }}\forall n \in {{\cal N}}.
\end{array}
    \end{equation}
 Now, with the help of ${\rm{C}}25'$, ${\rm{C}}26'$, ${\rm{C}}27'$ and ${\rm{C}}28'$, (45) can be recast into solving a sequence of convex optimization subproblem:

\begin{equation}
 \begin{array}{l}
({\rm{P8}})\mathop {\max }\limits_{\scriptstyle\;\;\;\;{\bf{W}} \succeq {\bf{0}},{{\bf{W}}_B} \succeq {\bf{0}},{\bf{Q}} \succeq {\bf{0}},\hfill\atop
\scriptstyle{\rm{      }}\varphi _{}^{worst},R_C^{worst},a^{worst},r^{worst}\hfill} {\rm{  }}\alpha \left( {\varphi _{}^{worst} - {\beta ^{worst}}} \right) + \left( {1 - \alpha } \right)R_C^{worst}\\
{\rm{\;\;\;\;\;\;\;\;\;\;\;\;s}}{\rm{.t}}{\rm{. C1', C2', C24}},{\rm{ C25', C26', C27', C28'}}{\rm{.}}
\end{array}
 \end{equation}

Now, (52) is a semidefinite program (SDP) which is convex and can be solved by using interior point methods, or standard convex optimization tool, e.g., CVX. The detailed procedure is summarized in Algorithm 2.
\begin{algorithm}
\caption{Solving (40).}

\label{alg1}
\begin{algorithmic}[1]
\STATE \textbf{Initialization:}

$\bullet$ Setting the initial values ${\bf{W}}_{}^0$, , ${\bf{W}}_B^0$, ${\bf{Q}}_{}^0$, ${\eta _0}$ and ${\eta _n}, n \in {{\cal N}}$.

$\bullet$ Setting maximum number of iterations $J_{}^{\max }$, iteration counter $j=1$ and iteration termination accuracy $\psi  > 0$.
\STATE \textbf{Repeat}
 \STATE Solving (52) with current ${\eta _0}$,  ${\eta _n}, n \in {{\cal N}}$ and obtaining the optimal ${\bf{W}}_{}^{\rm{*}}$, ${\bf{W}}_B^{\rm{*}}$ and ${\bf{Q}}_{}^{\rm{*}}$.
 \IF {${\left\|\! {{\bf{W}}_{}^{\rm{*}} \!- \!{\bf{W}}_{}^0} \!\right\|_2}\! +\! {\left\|\! {{\bf{W}}_B^{\rm{*}}\! -\! {\bf{W}}_B^0} \!\right\|_2} \!+ \!{\left\|\! {{\bf{Q}}_{}^{\bf{*}}{\bf{ - Q}}_{}^{\bf{0}}}\! \right\|_2} \!\le \!\psi$.}
      \STATE The optimal solutions of (56) are ${\bf{W}}_{}^{\rm{*}}$, ${\bf{W}}_B^{\rm{*}}$ and ${\bf{Q}}_{}^{\rm{*}}$.
      \STATE \textbf{Break}
 \ELSE
      \STATE Setting ${\eta _0} = \sqrt {\frac{{{\mathop{\rm Tr}\nolimits} \left( {{\bf{Q}}_{}^*{\bf{H}}_{BC}^{}} \right) + {\mathop{\rm Tr}\nolimits} \left( {{\bf{H}}_C^{}{\bf{W}}_{}^*} \right) + 1}}{{r_{}^{worst*}}}}$, ${\eta _n} = \sqrt {\frac{{{\mathop{\rm Tr}\nolimits} \left( {{\bf{H}}_{B,n}^{}{\bf{W}}_B^*} \right) + {\mathop{\rm Tr}\nolimits} \left( {{\bf{Q}}_{}^{\rm{*}}{\bf{H}}_{B,n}^{}} \right) + 1}}{{a_{}^{worst*}}}} ,\forall n \in {{\cal N}}$, ${\bf{W}}_{}^0 = {\bf{W}}_{}^*$, ${\bf{W}}_B^0 = {\bf{W}}_B^*$ and ${\bf{Q}}_{}^0 = {\bf{Q}}_{}^*$.
 \ENDIF
 \STATE $j=j+1$
 \STATE \textbf{Until} $j>J_{}^{\max }$
\end{algorithmic}
\end{algorithm}
\subsection{The Outer-Level Optimization }

 The outer-level is a single-variable optimization with respect to ${\beta ^{worst}}$. Let ${\rm{Y}}\left( {{\beta ^{worst}}} \right)$ be the optimized value of (52) with a given ${\beta ^{worst}}$, then the outer-level optimization is $\mathop {{\rm{max}}}\limits_{{\beta ^{worst}}} {\rm{Y}}\left( {{\beta ^{worst}}} \right)$. Note that
\begin{equation}
 \begin{array}{l}
0 \le {\beta ^{worst}} \le \mathop {\min }\limits_{n \in {{\cal N}}} \log \left( {1 + \frac{{{\mathop{\rm Tr}\nolimits} \left( {{\bf{H}}_{D,n}^{}{\bf{W}}} \right)}}{{{\mathop{\rm Tr}\nolimits} \left( {{\bf{H}}_{B,n}^{}{{\bf{W}}_B}} \right) + {\mathop{\rm Tr}\nolimits} \left( {{\bf{QH}}_{B,n}^{}} \right) + 1}}} \right) \le \\
\mathop {\min }\limits_{n \in {{\cal N}}} \log \left( {1 + {\mathop{\rm Tr}\nolimits} \left( {{\bf{H}}_{D,n}^{}{\bf{W}}} \right)} \right) \le \mathop {\min }\limits_{n \in {{\cal N}}} \log \left( {1 + MP_{\max }^{}\left\| {{\bf{h}}_{D,n}^{}} \right\|_2^2} \right).
\end{array}
  \end{equation}
 where the first inequality is due to the non-negative property of rate, the second inequality is because the secrecy rate is non-negative, and the last inequality comes from the fact that ${\mathop{\rm Tr}\nolimits} \left( {{\bf{H}}_{D,n}^{}{\bf{W}}} \right) \le {\mathop{\rm Tr}\nolimits} \left( {\bf{W}} \right)\left\| {{\bf{h}}_{D,n}^{}} \right\|_2^2$ for any ${\bf{W}} \succeq  {\bf{0}}$ and ${\bf{W}}[m] \le {P_{\max }}$. From (53), we can see that the optimization variable of the outer-level optimization has a bounded interval, and then we use one-dimensional search technology to solve the outer-level optimization.

 So far, we have solved ${\rm{P4\_1}}$, the approach to overcome ${\rm{P4\_2}}$ is similar to that of ${\rm{P4\_1}}$, except for replacing ${\rm{ C7'}}$ with ${\rm{ C7'\_2}}$ and we omit the details here.

{\color{blue}The whole process of solving the WRMP in the worst case is that, we first solve the ${\rm{P4\_1}}$ and test whether (39) is satisfied. If (39) is satisfied,the solution of ${\rm{P4\_1}}$ is the solution of the WRMP in the worst case, otherwise, we solve ${\rm{P4\_2}}$ and set the solution of it as the solution of the WRMP in the worst case.}
 \begin{remark}
{\color{blue} In our thousands of simulations results, the solution of (13) makes constraints C5 equal and the solution of ${\rm{P4\_1}}$ does not satisfy (39) never happen, which means eavesdroppers can not decode $s_B$ with large probability, even we do not deliberately set constraints to ensure this. In other words, the constraint C5 in (13) and ${\rm{P4\_2}}$ are redundant in most cases.}
\end{remark}

 \begin{remark}
{\color{blue} The reason why we do not consider (39) into the WRMP in the worst case is because (39) and $\rm C7'$ have the same structure, for overcoming $\rm C7'$, we deploy one-dimensional search technology, which means for jointly solving (39) and $\rm C7'$    a time-consuming two-dimensional search technology should be used. Also, as stated in remark 2, even we omit (39), the solution of ${\rm{P4\_1}}$ still satisfy (39) with large probability.}
\end{remark}
 \subsection{Rank-Profile Analysis}
  Note that, when the optimal solution $\left( {{\bf{W}}_{}^ * ,{\bf{W}}_B^ * } \right)$ of (44) satisfies ${\rm{Rank}}\left( {{\bf{W}}_{}^ * } \right) \le 1$ and ${\rm{Rank}}\left( {{\bf{W}}_B^ * } \right) \le 1$ the corresponding weighted maximum rate, i.e., $\alpha R_S^{worst} + \left( {1 - \alpha } \right)R_C^{worst}$ could be attained via single-stream transmit beamforming, which facilitates the implementation of physically realizable transceiver with low complexity. Also the condition ${\rm{Rank}}\left( {{\bf{W}}_{}^ * } \right) \le 1$ guarantees the relaxation from ${\rm{C}}7''$ to C24 is tight, i.e., C24 equivalent to ${\rm{C}}7''$ and (44) equivalent to (40). However, the rank constraint is non-convex and we omit it in (44), hence the optimal solution of (44) does not always lead to the optimal solution of the original problem (40). Apparently, if ${\bf{W}}_{}^ *$ (${\bf{W}}_B^ * $) is of rank one, then via eigenvalue decomposition, the optimal ${\bf{w}}_{}^ * $ (${\bf{w}}_B^ * $) of (40) can be easily obtained. Otherwise, some rank-one approximation procedures for ${\bf{W}}_{}^ *$ (${\bf{W}}_B^ * $) such as Gaussian randomization \cite{5447068} are applied to obtain a suboptimal solution of (40).

 The investigation on rank properties of ${\bf{W}}_{}^ * $, ${\bf{W}}_B^ * $ and ${{\bf{Q}}^{\bf{*}}}$ is still an open issue. However thankfully, we can prove that the rank one property holds for ${\bf{W}}_{}^ *$ in a special case.

\begin{proposition}

If there exist only one $\forall n \in {{\cal N}}$ satisfies $\log \left(\! {1 \!+\! \frac{{{\mathop{\rm Tr}\nolimits} \left(\! {{\bf{H}}_{D,n}^{}{\bf{W}}_{}^ * }\! \right)}}{{{\mathop{\rm Tr}\nolimits} \left(\! {{\bf{H}}_{B,n}^{}{\bf{W}}_B^{\bf{*}}} \!\right) \!+ \!{\mathop{\rm Tr}\nolimits} \left( \!{{\bf{QH}}_{B,n}^{}} \!\right) \!+ \!1}}} \!\right)\! - \!{\varphi ^{worst * }}{\rm{ \!= \!}}0$, then ${\bf{W}}_{}^ *$ must be of rank one.

$proof:$ Plase see Appendix A.
\end{proposition}

It is worth noting that, for D2D receivers it is almost impossible that, ${C_n} = {C_m},{\rm{ }}n,m \in {{\cal N}},n \ne m$ with similar $\left( {{\bf{W}}_{}^ * ,{\bf{W}}_B^ * ,{{\bf{Q}}^ * }} \right)$, since the channel conditions of each receiver are independent with each other, which implies that ${\rm{Rank}}\left( {{\bf{W}}_{}^ * } \right) = 1$ holds with large probability. In fact, in our thousands of simulations results, ${\rm{Rank}}\left( {{\bf{W}}_{}^ * } \right)>1$ never happens.

\section{Suboptimal Algorithms for High Dynamic Networks   }
In the above Section III and Section IV, we give the algorithms to obtain the optimal beamforming solutions for both the normal case and the worst case. However, the above algorithms need lots of auxiliary variables and contain a number of constraints, which may be time consuming and not suitable for high dynamic networks.

In this section, we propose a low complexity scheme to obtain a suboptimal beamforming solution. Let ${{\bf{G}}_{EC}} \buildrel \Delta \over = \left[ {{\bf{h}}_1^{},{\bf{h}}_2^{}, \ldots ,{\bf{h}}_L^{},{\bf{h}}_C^{}} \right]$ be the channel matrix from D2D transmitter group to eavesdropper group and CUE, the proposed scheme falls into two cases:$1) M>{\rm{Rank}}\left( {{{\bf{G}}_{EC}}} \right); 2) M<={\rm{Rank}}\left( {{{\bf{G}}_{EC}}} \right)$.

We note that although the proposed suboptimal solutions in this section are designed for the worst case, they could be easily adjusted for the normal case.

\subsection{Case One: $M>{\rm{Rank}}\left( {{{\bf{G}}_{EC}}} \right)$  }
In this case, D2D transmitter group directly put the confidential information onto the null space of the column space of ${{\bf{G}}_{EC}}$, i.e., no information is leaked to all eavesdroppers or interfered with CUE. In this case, ${\bf{w}}$ can be written as ${\bf{w}} = {\bf{G}}_{EC}^ \bot {\bf{\hat w}}$, where ${\bf{G}}_{EC}^ \bot  \in {\rm{\mathbb{C} }}{^{M \times (M - L - 1)}}$ is the standard orthogonal basis of the solution space of the equation ${\bf{G}}_{EC}^H{\bf{w}} = {\bf{0}}$ and ${\bf{\hat w}} \in{\rm{\mathbb{C} }} {^{(M - L - 1) \times 1}}$ is the new concerned optimization variable. It is easy to see ${{\bf{G}}_{EC}}{\bf{w}} = {{\bf{G}}_{EC}}{\bf{G}}_{EC}^ \bot {\bf{\hat w}}{\rm{ = }}{\bf{0}}$.

For BS, it adopts the MRT beamforming to focus the signal ${s_B}$ on CUE, and puts ${s_B}$ onto the null space of the column space of ${{\bf{G}}_{BN}}$, where ${{\bf{G}}_{BN}} \buildrel \Delta \over = \left[ {{\bf{h}}_{B,1}^{},{\bf{h}}_{B,2}^{}, \ldots ,{\bf{h}}_{B,N}^{}} \right]$ is the channel matrix from BS to D2D receiver group. In this case the beamforming vector of BS is
\begin{equation}
{\bf{w}}_B^{subopt}\! =\! \frac{{\left( {{\bf{I}}\! -\! {{\bf{G}}_{BN}}{{\left( {{\bf{G}}_{BN}^H{{\bf{G}}_{BN}}} \right)}^{ - 1}}{\bf{G}}_{BN}^H} \right){{\bf{h}}_{BC}}}}{{\left\| {\left( {{\bf{I}} \!-\! {{\bf{G}}_{BN}}{{\left( {{\bf{G}}_{BN}^H{{\bf{G}}_{BN}}} \right)}^{ - 1}}{\bf{G}}_{BN}^H} \right){{\bf{h}}_{BC}}} \right\|}}\sqrt {{P_B}}
\end{equation}
where $\left( {{\bf{I}} - {{\bf{G}}_{BN}}{{\left( {{\bf{G}}_{BN}^H{{\bf{G}}_{BN}}} \right)}^{ - 1}}{\bf{G}}_{BN}^H} \right)$ is the orthogonal projection matrix onto the null space of the column space of ${{\bf{G}}_{BN}}$. Since the transmitted signal, ${s_B}$, does not interfere the D2D receivers, the BS can use the maximal power to transmit ${s_B}$, i.e., the $\sqrt {{P_B}}$ term in (60).

Since no information is leaked to eavesdroppers, the original problem could be rewritten as
\begin{equation}
\begin{array}{l}
\mathop {\max }\limits_{{\bf{\hat w}},R_D^{subopt}} {\rm{  }}\alpha R_D^{subopt} + \left( {1 - \alpha } \right)R_C^{subopt}\\
{\rm{   s}}{\rm{.t}}{\rm{.  C29: }}{\left\| {{\bf{\hat w}}} \right\|^2} \le {P_{max}}\\
\;\;\;\;\;{\rm{C30: }}\log \left( {1 + {{\left| {{\bf{h}}_{D,n}^H{\bf{G}}_{EC}^ \bot {\bf{\hat w}}} \right|}^2}} \right) - R_D^{subopt} \ge 0,{\rm{ }}\forall n \in {{\cal N}}
\end{array}
\end{equation}
where $R_C^{subopt}{\rm{ = }}\log \left( {1 + {{\left| {{\bf{h}}_{BC}^H{\bf{w}}_B^{subopt}} \right|}^2}} \right)$ is the achievable rate of CUE. Since ${\bf{w}}_B^{subopt}$ is directly given by (60), we actually only optimize ${\bf{\hat w}}$ in (61), hence (61) can be further reduced to

\begin{equation}
\begin{array}{l}
\mathop {\max }\limits_{{\bf{\hat w}}} {\rm{  }}\mathop {{\rm{min}}}\limits_{\forall n \in {{\cal N}}} {\rm{ }}{\left| {{\bf{h}}_{D,n}^H{\bf{G}}_{EC}^ \bot {\bf{\hat w}}} \right|^2}\\
{\rm{   s}}{\rm{.t}}{\rm{.  C29: }}{\left\| {{\bf{\hat w}}} \right\|^2} \le {P_{max}}{\rm{        }}
\end{array}
\end{equation}
which is a simple max-min problem and can be solved by introducing an auxiliary variable $\kappa$ and then rewrite (56) as
\begin{equation}
\begin{array}{*{20}{l}}
\;\;\;{\mathop {\max }\limits_{{\bf{\hat w}},\kappa } {\rm{       }}\kappa }\\
\begin{array}{l}
{\rm{s}}.{\rm{t}}.{\rm{C29}}:{\left\| {{\bf{\hat w}}} \right\|^2} \le {P_{max}}\\
\;\;\;\;\;{\rm{C31}}:{\left| {{\bf{h}}_{D,n}^H{\bf{G}}_{EC}^ \bot {\bf{\hat w}}} \right|^2} \ge \kappa ,\forall n \in {{\cal N}}
\end{array}
\end{array}
\end{equation}
It is easy to see that, C31 can be approximated to convex form with the same approach of C11.
\subsection{Case Tow: $M\le {\rm{Rank}}\left( {{{\bf{G}}_{EC}}} \right)$  }
In this case, D2D transmitter group can hardly perform null space beamforming directly to prevent eavesdroppers from overhearing, since there has no non-zero solution of ${\bf{G}}_{EC}^H{\bf{x}} = {\bf{0}}$ when $M\le {\rm{Rank}}\left( {{{\bf{G}}_{EC}}} \right)$. Thus in this case the D2D transmitter group can hardly communication with D2D receiver group safely without the assistance of BS.

Based on the generalized singular value decomposition (GSVD), in this subsection we propose a BS-assisted scheme to guarantee that the confidential information wiretapped by Eve is protected by the jamming signal sent by BS.

Let ${\bf{w}}_B^i$ be the beamforming vector of BS for transmitting ${s_B} \sim CN(0,1)$ desired by CUE. Like Section V.A, BS puts the signal onto the null space of the column space of ${{\bf{G}}_{BNE}}{\rm{ = }}\left[ {{{\bf{G}}_{BN}},{{\bf{G}}_B}} \right]$, i.e., the transmitted ${s_B}$ does not interfere with either D2D receivers or Eve. Then ${\bf{w}}_B^i$ can be written as
\begin{equation}
\begin{array}{l}
{\bf{w}}_B^i = \\
\frac{{\left( {{\bf{I}} - {{\bf{G}}_{BNE}}{{\left( {{\bf{G}}_{BNE}^H{{\bf{G}}_{BNE}}} \right)}^{ - 1}}{\bf{G}}_{BNE}^H} \right){{\bf{h}}_{BC}}}}{{\left\| {\left( {{\bf{I}} - {{\bf{G}}_{BNE}}{{\left( {{\bf{G}}_{BNE}^H{{\bf{G}}_{BNE}}} \right)}^{ - 1}}{\bf{G}}_{BNE}^H} \right){{\bf{h}}_{BC}}} \right\|}}\sqrt {\left( {1 - \theta } \right){P_B}}
\end{array}
\end{equation}
where $\theta  \in \left[ {0,1} \right]$ is the power split factor.

Let ${\bf{w}}_B^{AN}$ be the beamforming vector of BS for transmitting jamming signal $n_B^{} \sim CN(0,1)$. Since the goal of the jamming signal is to interfere with Eve, ${\bf{w}}_B^{AN}$ should be orthogonal to the channel vectors from BS to D2D receivers and CUE. Let the channel matrix from BS to D2D receivers and CUE be ${{\bf{G}}_{BN{\rm{C}}}}{\rm{ = }}\left[ {{{\bf{G}}_{BN}},{\bf{h}}_{BC}^{}} \right]$, in order for ${\bf{w}}_B^{AN}$ to be orthogonal to the column space of ${{\bf{G}}_{BN{\rm{C}}}}$, ${\bf{w}}_B^{AN}$ is designed as ${\bf{w}}_B^{AN}{\rm{ = }}\frac{{{\bf{G}}_{BNC}^ \bot {\bf{\hat w}}_B^{AN}}}{{\left\| {{\bf{G}}_{BNC}^ \bot {\bf{\hat w}}_B^{AN}} \right\|}}\sqrt {\theta {P_B}}$, where $ {\bf{G}}_{BNC}^ \bot  \in {\rm{\mathbb{C} }}{^{K \times (K - N - 1)}}$ is the standard orthogonal basis of the solution space of the equation ${{\bf{G}}_{BN{\rm{C}}}}{\bf{w}}_B^{AN} = 0$ and ${\bf{\hat w}}_B^{AN} \in {\rm{\mathbb{C} }}{^{(K - N - 1) \times 1}}$ is the new concerned optimization variable.

For the D2D transmitter group, although when $M<=L+1$, the transmitted confidential information ${s_D}$ is bound to be overheard by Eve, we can prevent it from interfering CUE, i.e., ${\bf{w}}{\rm{ = }}\frac{{{\bf{h}}_C^ \bot {\bf{\hat w}}}}{{\left\| {{\bf{h}}_C^ \bot {\bf{\hat w}}} \right\|}}\sqrt {{P_{\max }}} $, where ${\bf{h}}_C^ \bot  \in\mathbb{C} {^{M \times (M - 1)}}$ is the standard orthogonal basis of the solution space of the equation ${\bf{h}}_C^H{\bf{w}} = {\bf{0}}$ and ${\bf{\hat w}} \in\mathbb{C} {^{M - 1}}$ is the new concerned optimization variable.

Based on the above discussion, the received signal of Eve can be written as

\begin{equation}
\begin{array}{l}
{{\bf{y}}_E} = {\bf{G}}_E^H\frac{{{\bf{h}}_C^ \bot {\bf{\hat w}}}}{{\left\| {{\bf{h}}_C^ \bot {\bf{\hat w}}} \right\|}}\sqrt {{P_{tot}}} {s_D}\\
\; \; \; \;\;  \; + {\bf{G}}_B^H\frac{{{\bf{G}}_{BNC}^ \bot {\bf{\hat w}}_B^{AN}}}{{\left\| {{\bf{G}}_{BNC}^ \bot {\bf{\hat w}}_B^{AN}} \right\|}}\sqrt {\theta {P_B}} {n_B} + {{\bf{n}}_E}.
\end{array}
\end{equation}

Define ${\bf{{\rm E}}}{\rm{ = }}{\bf{G}}_E^H{\bf{h}}_C^ \bot $ and ${\bf{F}} = {\bf{G}}_B^H{\bf{G}}_{BNC}^ \bot$ as the equivalent channel matrixes of the eavesdrop channel from D2D transmitter group to Eve and the jamming channel from BS to Eve, respectively. For guaranteeing that the confidential information wiretapped by Eve are protected by the jamming signal sent by BS, i.e., Eve can hardly separate ${s_D}$ from ${{\bf{y}}_E}$, we need ${\mathop{\rm span}\nolimits} \left( {{\bf{E\hat w}}} \right){\rm{ = span}}\left( {{\bf{F\hat w}}_B^{AN}} \right)$, which means the confidential information signal ${s_D}$ is always mixed with the jamming signal ${n_B}$. To handle the alignment constraint, GSVD, which can yield the decoupled parallel sub channels, is applied as follow:
\begin{equation}
\left[ {{\bf{U,V,X,R,S}}} \right] = {\mathop{\rm gsvd}\nolimits} \left( {{\bf{F}}_{}^H{\rm{,}}{\bf{E}}_{}^H} \right)
\end{equation}
where ${\bf{U}} \in\mathbb{C} _{}^{(K - N - 1) \times (K - N - 1)}$ and ${\bf{V}} \in\mathbb{C} _{}^{(M - 1) \times (M - 1)}$ are unitary matrices, ${\bf{X}} \in\mathbb{C} {^{L \times L}}$, ${\bf{R}} \in\mathbb{C} _{}^{(K - N - 1) \times L}{\rm{ = }}\left[ \begin{array}{l}
{{\bf{\Lambda }}_1}\\
{\bf{0}}
\end{array} \right]$, and ${\bf{S}} \in\mathbb{C} _{}^{(M - 1) \times L}{\rm{ = }}\left[ {{{\bf{\Lambda }}_2},{\bf{0}}} \right]$, ${{\bf{\Lambda }}_1}{\rm{ = diag}}\left( {\Lambda _{1,1}^{},\Lambda _{1,2}^{}, \ldots \Lambda _{1,L}^{}} \right)$ is a diagonal matrix with ascending sort and ${{\bf{\Lambda }}_2}{\rm{ = diag}}\left( {\Lambda _{2,1}^{},\Lambda _{2,2}^{}, \ldots \Lambda _{2,M - 1}^{}} \right)$ is a diagonal matrix with descending sort. According to the definition of GSVD we obtain that ${\bf{F}} = {\bf{XR}}_{}^H{{\bf{U}}^H} = {\bf{X}}\left[ {{{\bf{\Lambda }}_1},{\bf{0}}} \right]{{\bf{U}}^H}$ and ${\bf{E}} = {\bf{XS}}_{}^H{{\bf{V}}^H} = {\bf{X}}\left[ \begin{array}{l}
{{\bf{\Lambda }}_2}\\
{\bf{0}}
\end{array} \right]{{\bf{V}}^H}$. Based on the results of GSVD, we can equivalently treat the equivalent eavesdrop channel and the equivalent jamming channel as a series of subchannels, respectively.

Let us first check the equivalent eavesdrop channel matrix ${\bf{E}}$. If we select the $x^{th}$ column of ${\bf{V}}$ as the beamforming vector ${\bf{\hat w}}$, then we have ${\bf{E\hat w}} = {\bf{X}}\left( {:,x} \right)\Lambda _{2,x}^{}$, and $\Lambda _{2,x}^{}$ can be treated as the equivalent channel coefficient of the $x^{th}$ subchannel. Because our goal is to prevent Eve from overhearing, hence, we select the $(M-1)^{th}$ column of ${\bf{V}}$, i.e., ${\bf{V}}\left( {:,M - 1} \right)$ as the beamforming vector ${\bf{\hat w}}$, which corresponds to the smallest equivalent channel coefficient of the equivalent eavesdrop channel ${\bf{E}}$.

For guaranteeing the confidential information signal ${s_D}$ is protected by the jamming signal ${n_B}$, we also choose the corresponding $(M-1)^{th}$ column of ${\bf{U}}$ as the beamforming vector ${\bf{\hat w}}_B^{AN}$, i.e., ${\bf{\hat w}}_B^{AN} = {\bf{U}}\left( {:,M - 1} \right)$.

Substituting the selected beamforming vectors ${\bf{\hat w}}$ and  ${\bf{\hat w}}_B^{AN}$ into (60), we have
\begin{equation}
\begin{array}{l}
{{\bf{y}}_E} = {\bf{X}}\left( {:,M - 1} \right)\Lambda _{2,M - 1}^{}\frac{{\sqrt {{P_{max}}} {s_D}}}{{\left\| {{\bf{h}}_C^ \bot {\bf{\hat w}}} \right\|}}\\
\;\;\;\;\;\;+ {\bf{X}}\left( {:,M - 1} \right)\Lambda _{1,M - 1}^{}\frac{{\sqrt {\theta {P_B}} {n_B}}}{{\left\| {{\bf{G}}_{BNC}^ \bot {\bf{\hat w}}_B^{AN}} \right\|}} + {{\bf{n}}_E}.
\end{array}
\end{equation}

Next, we move forward to calculate the optimal power split factor $\theta$. For simplicity, we observe that the interferences in (66) are mainly composed of the jamming signal ${{{n}}_B}$, and hence we ignore the AWGN term ${{\bf{n}}_E}$ in (66).  With the MMSE receiver, the eavesdrop channel capacity of Eve is
\begin{equation}
C_E^{subopt}{\rm{ = }}\log \left( {1{\rm{ + }}\frac{{\Lambda _{2,M - 1}^2{P_{max}}}}{{\Lambda _{1,M - 1}^2\theta {P_B}}}} \right)
\end{equation}

Now, the weighted rate sum becomes
\begin{equation}
\begin{array}{l}
\alpha \left( {\mathop {\rm{min}}\limits_{\forall n \in {{\cal N}}} \left( {\log \left( {1 + I{P_{max}}} \right)} \right) - \log \left( {1{\rm{ + }}\frac{{\Lambda _{2,M - 1}^2{P_{max}}}}{{\Lambda _{1,M - 1}^2\theta {P_B}}}} \right)} \right)\\
 + \left( {1 - \alpha } \right)\log \left( {1{\rm{ + }}J\left( {1 - \theta } \right){P_B}} \right)
\end{array}
\end{equation}
where
\begin{equation}
I{\rm{ = }}{\left| {{\bf{h}}_{D,n}^H\frac{{{\bf{h}}_C^ \bot {\bf{\hat w}}}}{{\left\| {{\bf{h}}_C^ \bot {\bf{\hat w}}} \right\|}}} \right|^2}
\end{equation}
and
\begin{equation}
J = {\left| {{\bf{h}}_{BC}^H\frac{{\left( {{\bf{I}} - {{\bf{G}}_{BNE}}{{\left( {{\bf{G}}_{BNE}^H{{\bf{G}}_{BNE}}} \right)}^{ - 1}}{\bf{G}}_{BNE}^H} \right){{\bf{h}}_{BC}}}}{{\left\| {\left( {{\bf{I}} - {{\bf{G}}_{BNE}}{{\left( {{\bf{G}}_{BNE}^H{{\bf{G}}_{BNE}}} \right)}^{ - 1}}{\bf{G}}_{BNE}^H} \right){{\bf{h}}_{BC}}} \right\|}}} \right|^2}.
\end{equation}

After the first order derivative, it is easy to obtain that the optimal $\theta $ to maximize (68) is
\begin{equation}
\theta _{}^ * {\rm{ = }}\min (\frac{{ - B - \sqrt {{B^2} - 4AC} }}{{2A}},1)
\end{equation}
in which
\begin{equation}
A =  - (1 - \alpha )J\Lambda _{1,M - 1}^4P_B^3
\end{equation}
\begin{equation}
B =  - \Lambda _{2,M - 1}^2{P_{max}}\Lambda _{1,M - 1}^2P_B^2J
\end{equation}
and
\begin{equation}
C = \alpha \Lambda _{2,M - 1}^2P_{\max }^{}\Lambda _{1,M - 1}^2P_B^2J + \alpha \Lambda _{1,M - 1}^2{P_B}\Lambda _{2,M - 1}^2P_{\max }^{}
\end{equation}
In the suboptimal solution, we do not consider the individual power constraint of each UE in VAA. If the individual power constraint is unsatisfied, the suboptimal solution should be scaled to satisfy the individual power constraint.
\section{Simulation Results}

This section presents simulation results to validate the performance of our proposed algorithms. The transmit powers are normalized with AWGN power ${N_0} = 1$. That is, the transmit power $P$ dB means the transmit power is ${10^{P/10}}$ times higher than the AWGN power. All channels are considered to be independent quasi-static block Rayleigh fading with ${\cal {CN}}(0,1)$\footnote{\color{blue}We do not consider the specific distribution of each node, and set all channels are rayleigh distribution to reflect the randomness distribution of each node in this paper. } . The simulation results are collected from Monte Carlo simulations with 100 independent channel realizations.

We first present the impact of the number of the D2D transmitter on the achievable rate. In Fig. 2, the number of eavesdroppers, $L$ is set to zero which corresponds to the normal D2DM communication scenario without the threat of eavesdroppers. Fig. 2 shows that, when $M=1$, D2DM can achieve only a certain rate through the power control of the transmitter and the BS. While when $M>1$, the rate of VD2DM increases dramatically, that is because multiple D2D transmitters could form a VAA and play cooperative beamforming to focus the signal on the D2D receiver who has the worst channel conditions. We note that the number of D2D transmitter $M=1$ represents the traditional D2DM without the assistance of VAA, hence the results demonstrate that the VAA-aided communication can significantly improve the efficiency of D2D multicast communication compare to the traditional D2DM.

We then show the secrecy rate obtained via the aided of VAA under the threat of multiple eavesdroppers. In Fig. 3, when $M=1$, D2DM can achieve a tiny secrecy rate by relying on the cooperative interference of the BS. However, when $M>1$, the secrecy rate of VD2DM increases dramatically, because the multiple D2D transmitters form a VAA and play cooperative beamforming to focus the signal on the D2D receivers while prevent the eavesdroppers to overhear. Similar to Fig. 2, the number of D2D transmitter $M=1$ represents the traditional secrecy D2DM communication without the help of VAA, hence the results in Fig. 3 validate the secrecy improvement brought by VAA.

\begin{figure}
\centering
\begin{minipage}[t]{0.48\textwidth}
\centering
\includegraphics[width=2.8in]{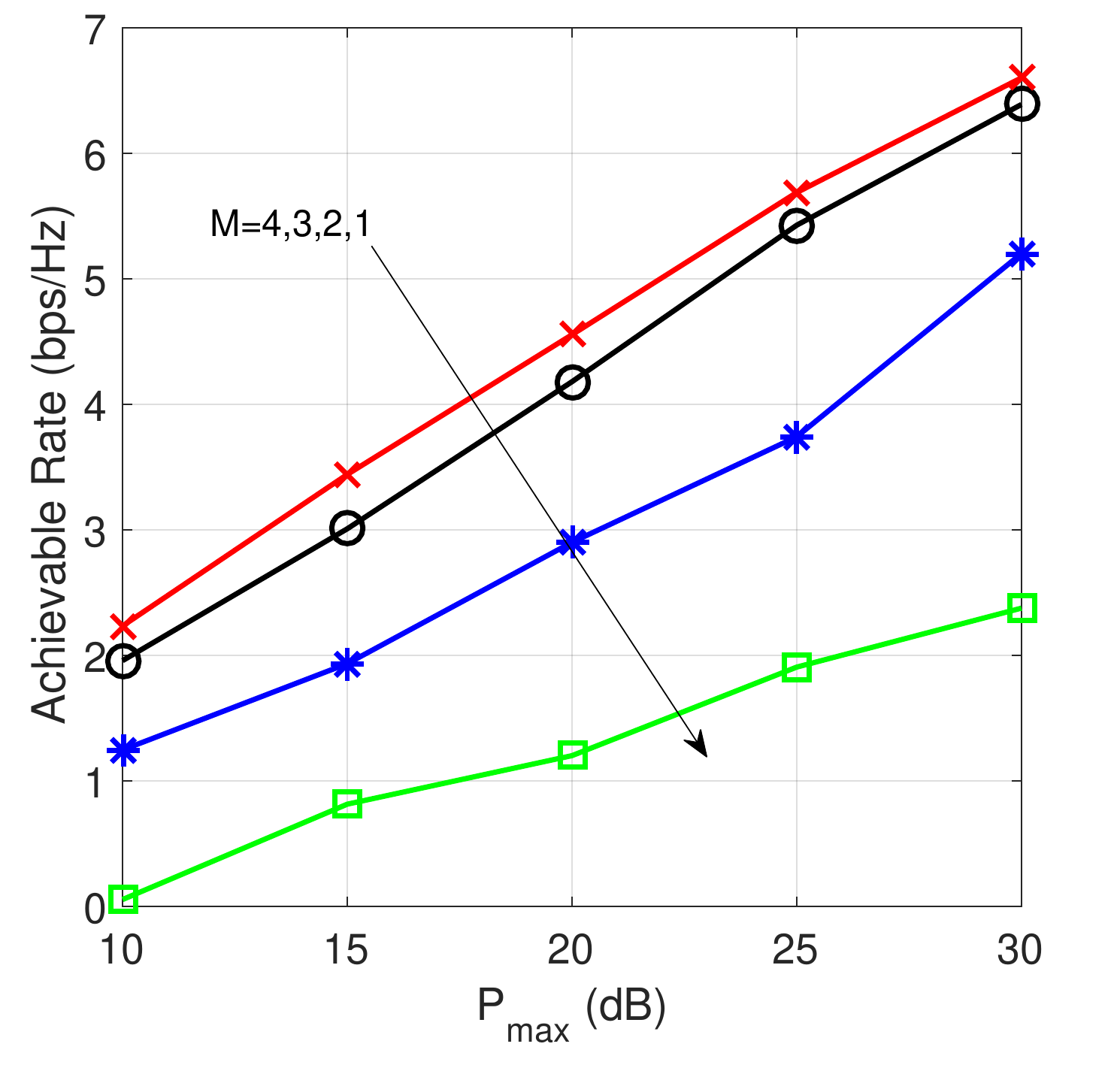}
\caption{The achievable rate of D2DM versus $P_{\rm{max}}$ with respect to different number of D2D transmitters, in which $K=15, N=5, L=0$ and $P_{B}=40$ dB.}
\end{minipage}
\begin{minipage}[t]{0.48\textwidth}
\centering
\includegraphics[width=3in]{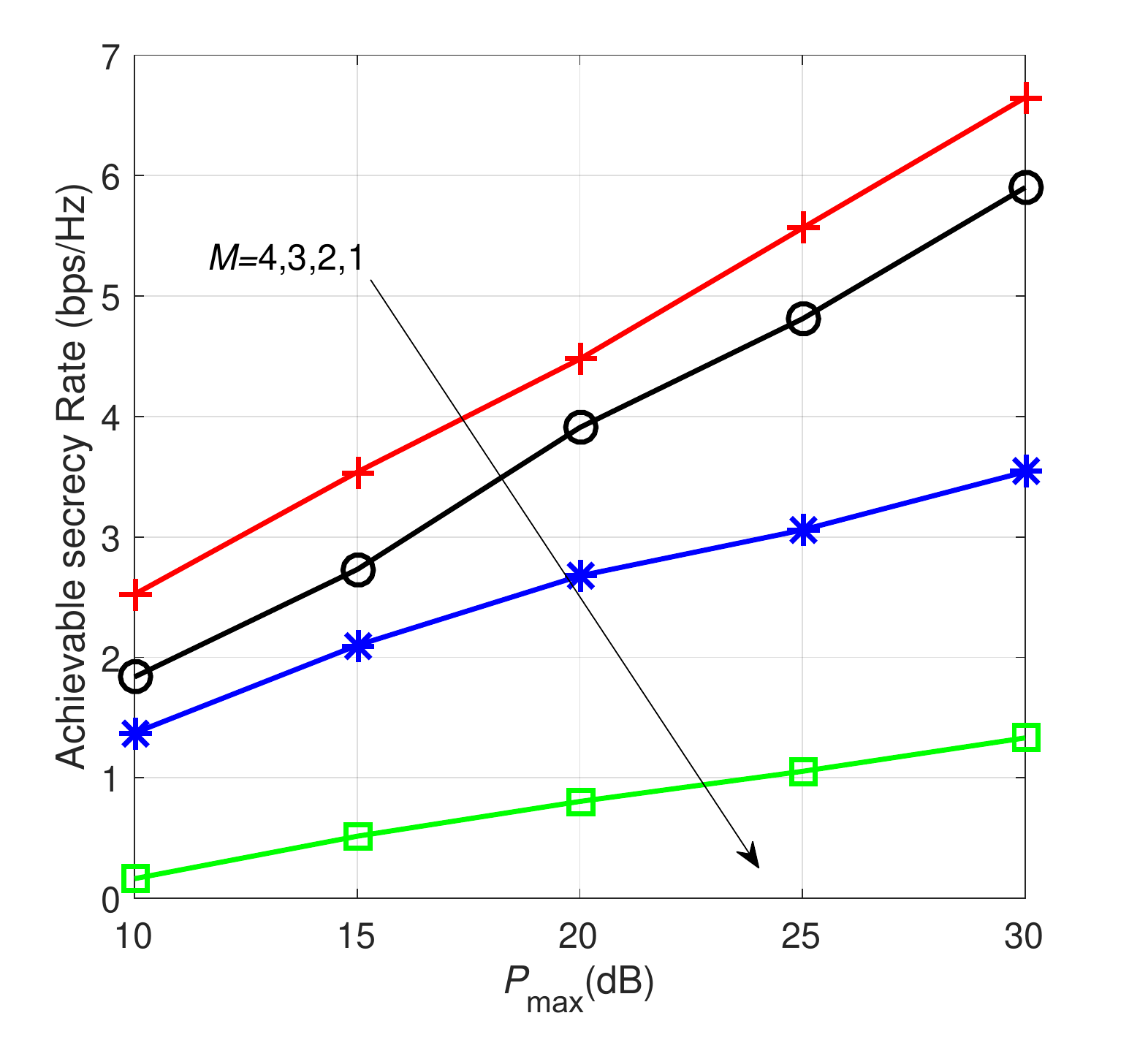}
\caption{The achievable secrecy rate of D2DM versus $P_{\rm{max}}$ with respect to different number of D2D transmitters, in which $K=15, N=5, L=5$ and $P_{B}=40$ dB.}
\end{minipage}
\end{figure}

We next present the achievable secrecy rate of VD2DM with respect to different maximum transmit power of D2D transmitter group. As can be seen in Fig. 4, the achievable secrecy rate of VD2DM monotonously increases with the increase of the maximum transmit power of D2D transmitter group. Also, a larger $\alpha $ will lead to a larger achievable secrecy rate. Further more, in the worst case, due to the strong overhearing capability provided by the receive VAA, the achievable secrecy rate is lower than that of the normal case. However, with the help of AN, the gap between the achievable secrecy rate of the worst case and that of the normal case is small, indicating that the proposed scheme can effectively secure the communication of VD2DM under the worst case. It is also shown in Fig.4 that the suboptimal secure solution proposed in Section V can provide a significant achievable secrecy rate, although it is not as good as the optimal solution.
\begin{figure}
\centering
\begin{minipage}[t]{0.48\textwidth}
\centering
\includegraphics[width=2.6in]{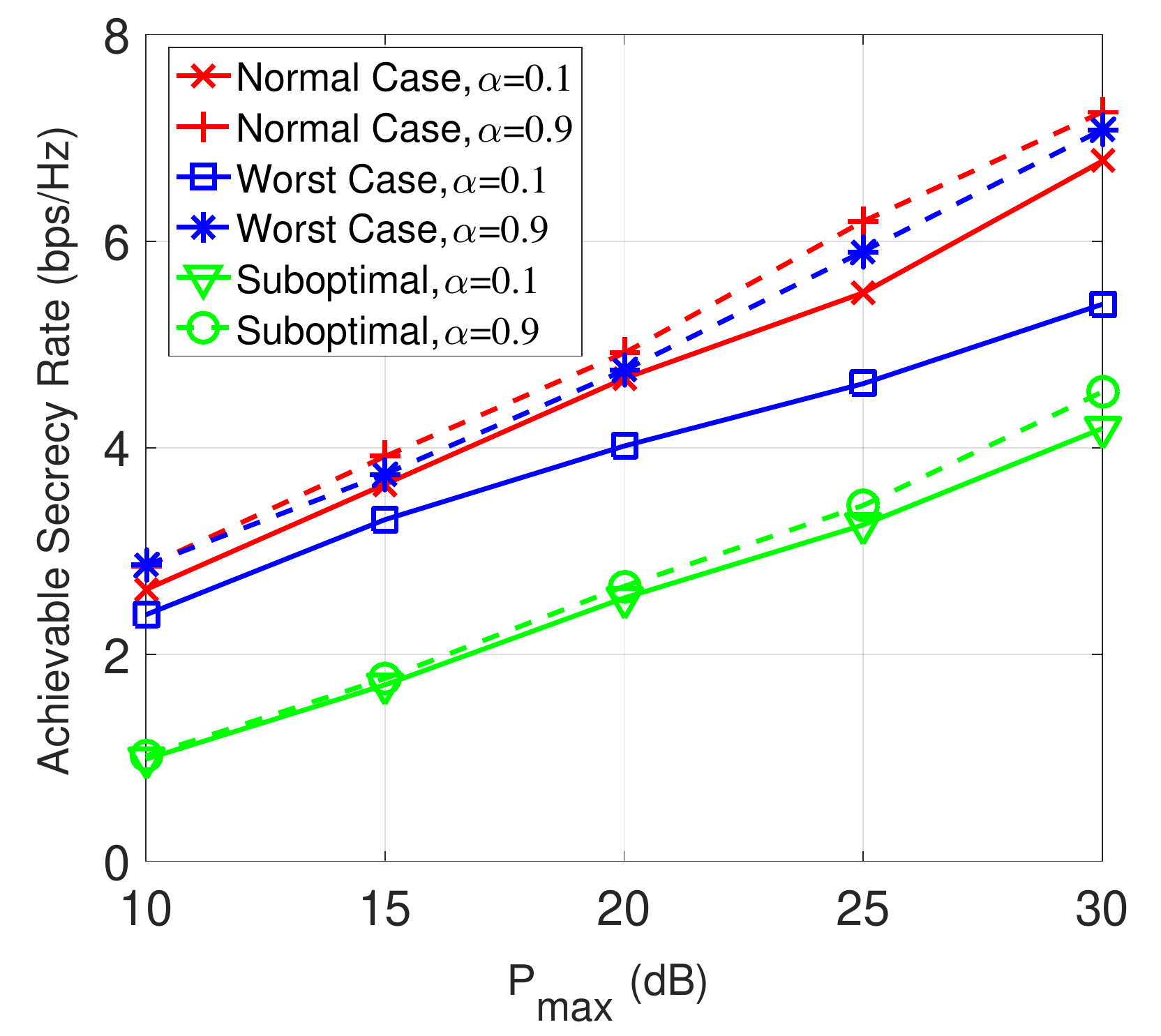}
\caption{The achievable secrecy rate of D2DM versus $P_{\rm{max}}$, in which $K=15, M=5, N=5, L=5$ and $P_{B}=40$ dB. }
\end{minipage}
\begin{minipage}[t]{0.48\textwidth}
\centering
\includegraphics[width=3in]{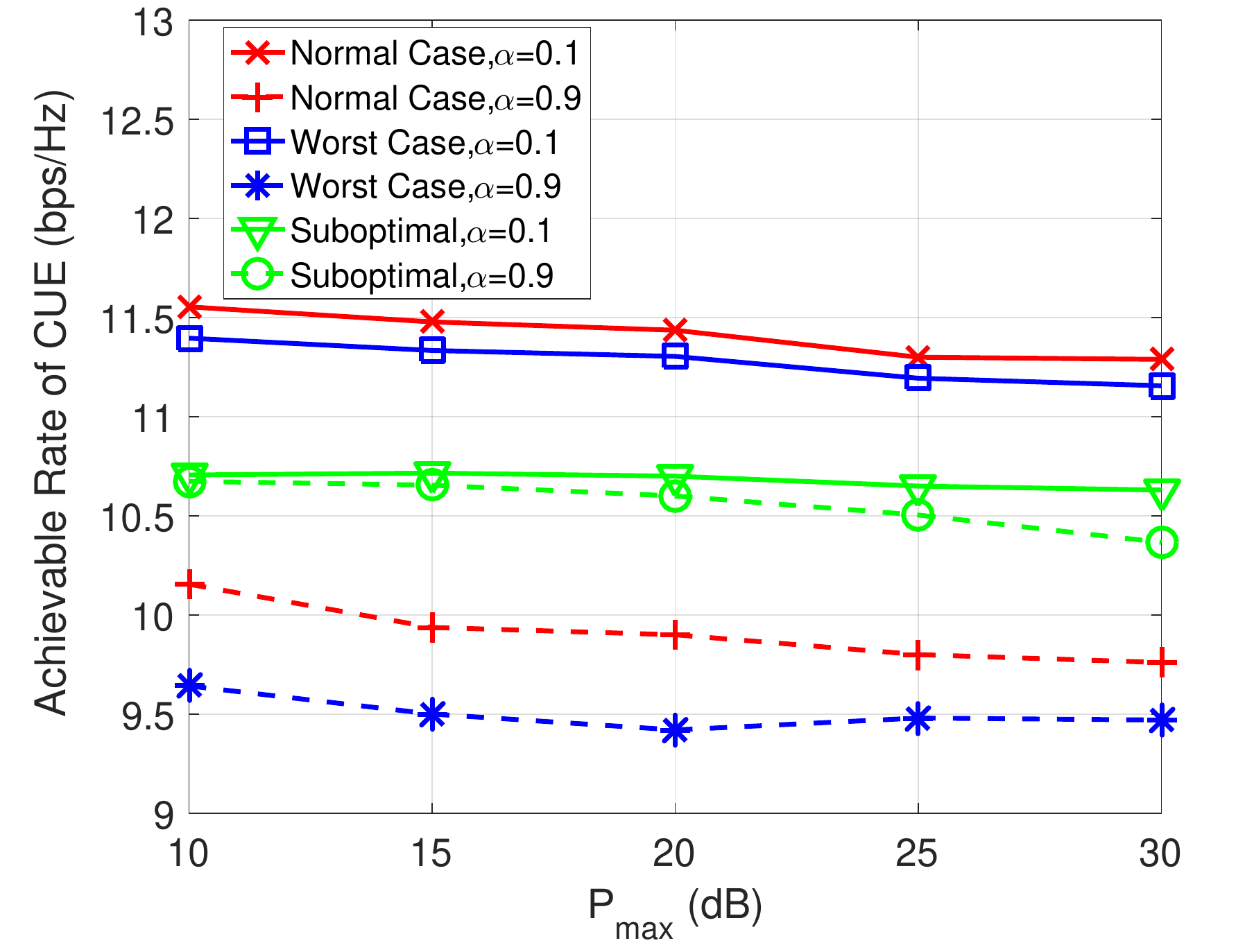}
\caption{The achievable rate of CUE with respect to $P_{\rm{max}}$, where $K=15, M=5, N=5, L=5$ and $P_{B}=40$ dB. }
\end{minipage}
\end{figure}

In Fig. 5 we show the achievable rate of CUE as a function of the maximum transmit power of D2D transmitter group with different $\alpha $. It is easy to see the maximum transmit power of D2D transmitter group has little effect on CUE's achievable rate. That is because the proposed algorithms in the normal case and the worst case are to jointly maximize the achievable secrecy rate of VD2DM and achievable rate of CUE, hence the interference at CUE caused by VD2DM keeps limited to a small level. Also, in the suboptimal algorithm, the null-space beamforming ensures that VD2DM does not interfere with CUE. On the other hand, the reason why the achievable rate of the worst case is lower than that of the normal case is because in the worst case, BS need to split some energy to transmit AN, which decreases the energy to transmit the signal CUE desired. {\color{blue} Another interesting observation shown in Fig. 5 is that, when $\alpha {\rm{ = }}0.9$, the achievable rate of the suboptimal solution is higher than that of both algorithm 1 and 2. That is because when $\alpha {\rm{ = }}0.9$, both algorithm 1 and 2 tend to achieve a higher secrecy rate of D2DM rather then a higher data rate of CUE, which yields more interference at CUE caused by D2D transmitters, By contrast, in the suboptimal scheme, the CUE is always under the null space of D2D transmitters, which means D2D transmitters never interfere with CUE and hence the rate of CUE of the suboptimal algorithm is higher than that of algorithm 1 and 2, note that, the weight sum rates of optimal algorithms are always higher than that of the suboptimal algorithm, whatever the value of $\alpha$ is. }

In Fig. 6 we present the impacts of the number of eavesdroppers, $L$, on the achievable secrecy rate. As the number of eavesdroppers increases, the achievable secrecy rate decreases. However, the decreasing trends of achievable secrecy rate are not obvious in both the normal case and the worst case. This is because the proposed algorithms can effectively limit the overhearing capacity of eavesdroppers by jointly optimizing the beamforming vector of the D2D transmitter group and BS. Another point of concern is that, for the suboptimal solution, when $L$ increases from 1 to 3, the decreasing trend of achievable secrecy rate is obvious, but when $L$ increases from 4 to 9, the decreasing trend is not obvious. At last, in the normal case and the worst case, the achievable data rates of CUE are almost unaffected by the number of eavesdroppers, since eavesdroppers only overhear VD2DM and the BS has enough power budget to transmit AN without affecting the transmission of $s_B$.
\begin{figure}
\centering
\begin{minipage}[t]{0.48\textwidth}
\centering
\includegraphics[width=2.8in]{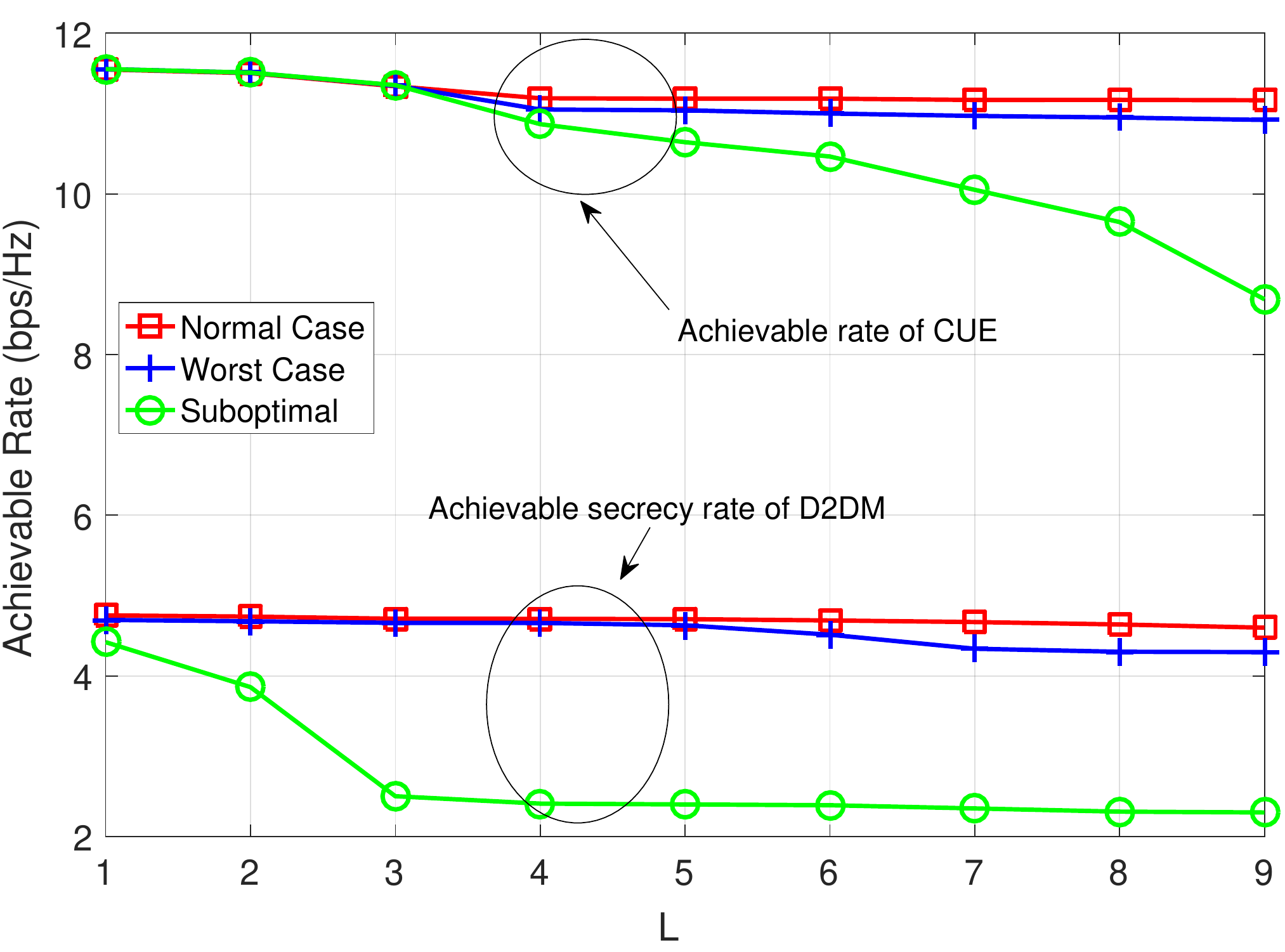}
\caption{Achievable rates with respect to different number of eavesdropper $L$, where $K=15, M=5, N=5$, $\alpha=0.5 $, ${P_{\max }} = 20$ dB and ${P_B} = 40$ dB. }
\end{minipage}
\begin{minipage}[t]{0.48\textwidth}
\centering
\includegraphics[width=2.8in]{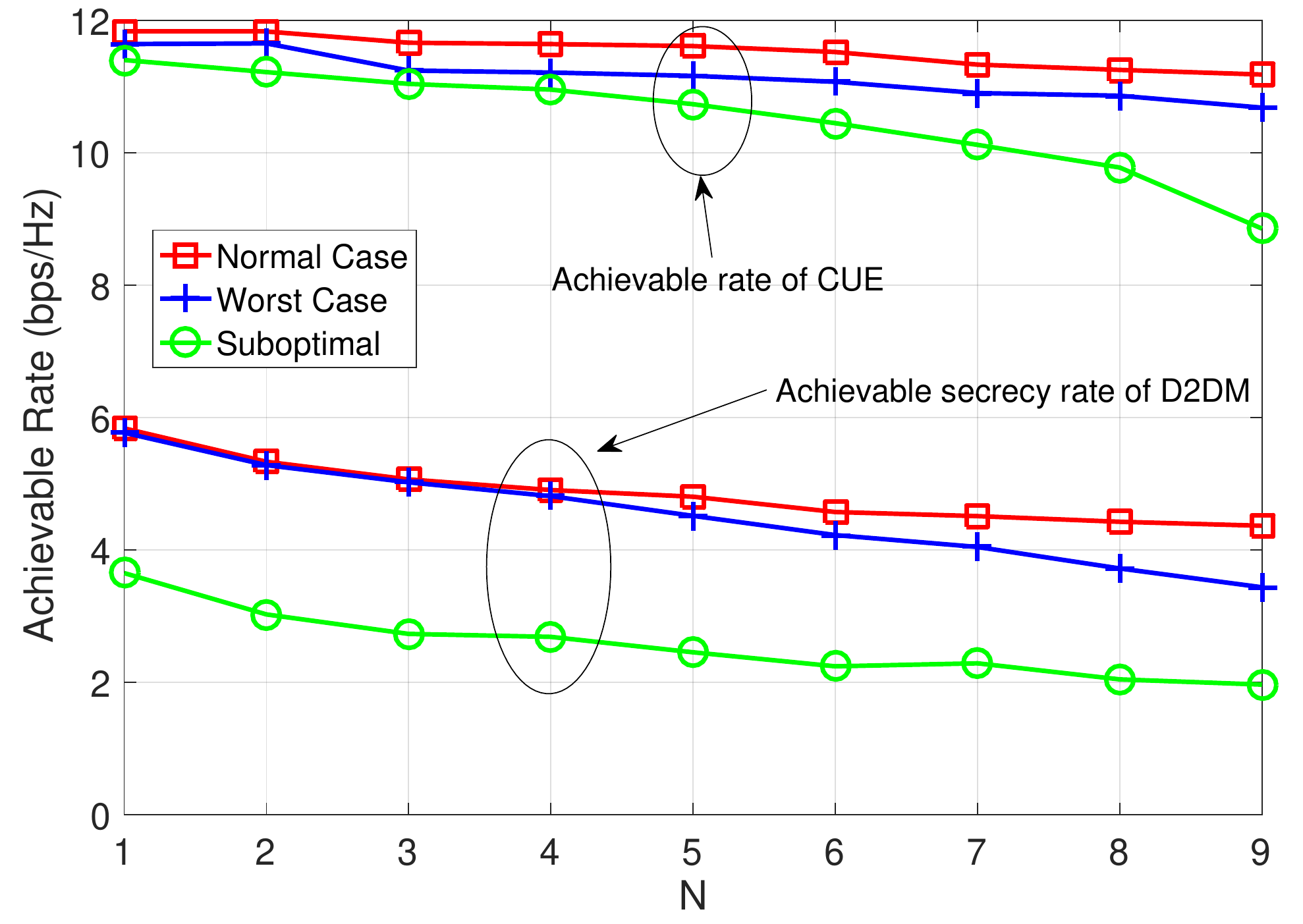}
\caption{Achievable rates with respect to different number of D2D receivers $N$, where $K=15, M=5, L=5$, $\alpha=0.5 $, ${P_{\max }} = 20$ dB and ${P_B} = 40$ dB. }
\end{minipage}
\begin{minipage}[t]{0.48\textwidth}
\centering
\includegraphics[width=2.8in]{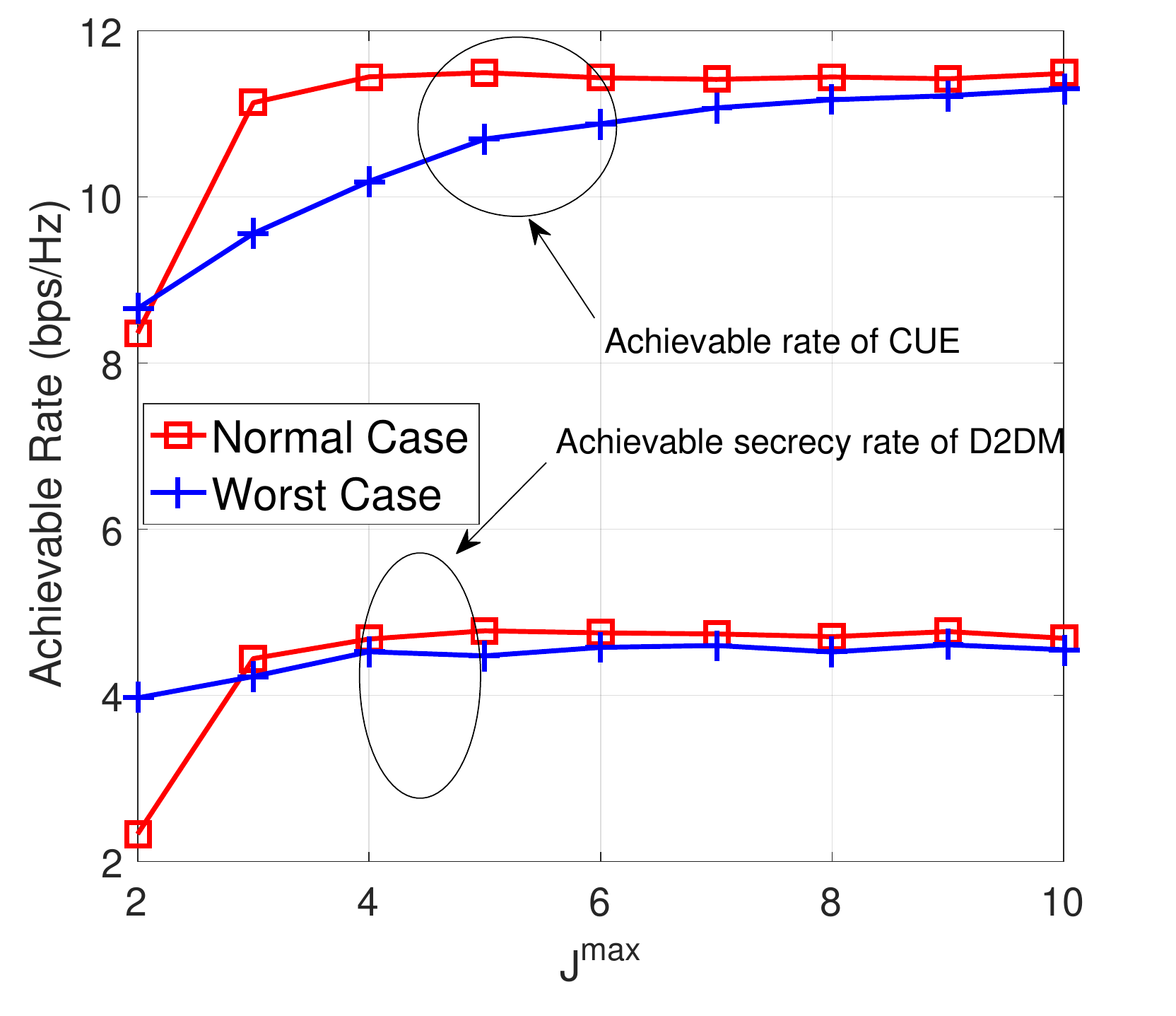}
\caption{ Convergence behavior for the achievable rates, where $K=15$, $M=5$, $N=5$, $L=5$, $\alpha=0.5 $, ${P_{\max }} = 20$ dB and ${P_B} = 40$ dB. }
\end{minipage}
\end{figure}

The relationship between achievable rates and the number of D2D receivers is illustrated in Fig. 7. We can see that as the number of D2D receivers increases, the achievable secrecy rate decreases slowly. The reason why more D2D receivers result in lower achievable secrecy rate is because the larger the number of D2D receivers, the higher the probability of a poor channel condition of the worst channel receiver. However, the proposed algorithms in both the normal case and the worst case are max-min based algorithms which take care of the worst channel receivers. Hence, the trend of decreasing is slowly.

As above mentioned, as the number of receivers increase, it becomes more difficult to ensure communication security. Thus, the proposed algorithms trend to use the BS's signals (such as signal desired by CUE and AN) to interfere with eavesdroppers for increasing the achievable secrecy rate, which makes the achievable rate of CUE decrease slowly.

At last, the overall convergence process of the proposed algorithms is shown in Fig. 8. It can be see that, setting the maximum iteration number of Algorithm 1 and 2 as five is enough to achieve the near optimal solution, which means the proposed algorithms can converge quickly.

\section{Conclusion}

This paper has studied the secure transmission design for a VAA-aided D2D multicast communication system under the threat of multiple eavesdroppers. By treating eavesdroppers as either independent ovearhearing or joint overhearing, respectively, two different non-convex optimization problems have been proposed and solved to achieve the sum data rate of both the cellular user equipment and
the achievable secrecy rate of the D2D multicast communication. Also a time-saving suboptimal beamforming scheme has been proposed to adapt to the high dynamic network. Simulation results
have validated the advantages of the VAA-aided D2D multicast scheme
from the perspective of wireless communication security.
  \begin{figure*}[t]
\vspace*{2pt}
\hrulefill 
\begin{small}
 \begin{equation}
\begin{array}{l}
\mathop {\min }\limits_{{\bf{W}} \succeq {\bf{0}},{{\bf{W}}_B} \succeq {\bf{0}},{\bf{Q}} \succeq {\bf{0}}} {\rm{  Tr}}\left( {{{\bf{W}}_{}}} \right) + {\rm{Tr}}\left( {{{\bf{W}}_B}} \right) + {\mathop{\rm Tr}\nolimits} \left( {\bf{Q}} \right)\\
{\rm{   s}}{\rm{.t}}{\rm{. C1',C2'}}\\
{\rm{         C}}4''':\log \left( {1 + \frac{{{\mathop{\rm Tr}\nolimits} \left( {{\bf{H}}_{BC}^{}{{\bf{W}}_B}} \right)}}{{{\mathop{\rm Tr}\nolimits} \left( {{\bf{QH}}_{BC}^{}} \right) + {\mathop{\rm Tr}\nolimits} \left( {{\bf{H}}_C^{}{\bf{W}}} \right) + 1}}} \right) \ge R_C^{worst*} \\
{ \Leftrightarrow {{\bf{e}}^{R_C^{worst*}}}\left( {{\rm{Tr}}\left( {{\bf{QH}}_{BC}^{}} \right) + {\rm{Tr}}\left( {{\bf{H}}_C^{}{\bf{W}}} \right) + 1} \right) - {\rm{Tr}}\left( {{\bf{H}}_{BC}^{}{{\bf{W}}_B} + {\bf{QH}}_{BC}^{} + {\bf{H}}_C^{}{\bf{W}}} \right) - 1 \le 0}\\
{\rm{         C6''': }}\log \left( {1 + \frac{{{\mathop{\rm Tr}\nolimits} \left( {{\bf{H}}_{D,n}^{}{\bf{W}}} \right)}}{{{\mathop{\rm Tr}\nolimits} \left( {{\bf{H}}_{B,n}^{}{{\bf{W}}_B}} \right) + {\mathop{\rm Tr}\nolimits} \left( {{\bf{QH}}_{B,n}^{}} \right) + 1}}} \right) - \varphi _{}^{worst*}  \ge 0\\
{ \Leftrightarrow {{\bf{e}}^{\varphi _{}^{worst*}}}\left( {{\rm{Tr}}\left( {{\bf{H}}_{B,n}^{}{{\bf{W}}_B}} \right) + {\rm{Tr}}\left( {{\bf{QH}}_{B,n}^{}} \right) + 1} \right) - {\rm{Tr}}\left( {{\bf{H}}_{B,n}^{}{{\bf{W}}_B} + {\bf{QH}}_{B,n}^{} + {\bf{H}}_{D,n}^{}{\bf{W}}} \right) - 1 \le 0,\forall n \in \cal{N}}\\
{\rm{         C24': }}\left( {{{\bf{e}}^{{\beta ^{worst*}}}} - 1} \right)\left( {{\bf{G}}_B^H{{\bf{W}}_B}{\bf{G}}_B^{} + {\bf{G}}_B^H{\bf{QG}}_B^{} + {{\bf{I}}_L}} \right) - {\bf{G}}_E^H{\bf{WG}}_E^{} \succeq {\bf{0}}.
\end{array}
 \end{equation}
 \end{small}
\end{figure*}
  \begin{figure*}[t]
\vspace*{2pt}
\hrulefill 
\begin{small}
 \begin{equation}
\begin{array}{l}
L\left( {{\bf{W}},{{\bf{W}}_B},{\bf{Q}},\varepsilon ,\lambda ,\mu ,{\nu _n},{\bf{A}},{\bf{B,C,D}}} \right) = \\
{\rm{Tr}}\left( {{{\bf{W}}_{}}} \right) + {\rm{Tr}}\left( {{{\bf{W}}_B}} \right)\! + \!{\mathop{\rm Tr}\nolimits} \left( \!{\bf{Q}}\! \right){\rm{\! +\! }}\sum\limits_{m = 1}^M {{\varepsilon _m}\left( {{\bf{W}}\left[ m \right] - {P_{\max }}/M} \right)} \! + \!\lambda \left(\! {{\rm{Tr}}\left(\! {{{\bf{W}}_B}}\! \right) \!+ \!{\mathop{\rm Tr}\nolimits} \left(\! {\bf{Q}} \!\right)\! - \!{P_B}}\! \right)\\
{ + \mu \left( {{{\bf{e}}^{R_C^{worst*}}}\left( {{\rm{Tr}}\left( {{\bf{QH}}_{BC}^{}} \right) + {\rm{Tr}}\left( {{\bf{H}}_C^{}{\bf{W}}} \right) + 1} \right) - {\rm{Tr}}\left( {{\bf{H}}_{BC}^{}{{\bf{W}}_B} + {\bf{QH}}_{BC}^{} + {\bf{H}}_C^{}{\bf{W}}} \right) - 1} \right)}\\
 { + \sum\limits_{n = 1}^N {{\nu _n}\left( {{{\bf{e}}^{\varphi _{}^{worst*}}}\left( {{\rm{Tr}}\left( {{\bf{H}}_{B,n}^{}{{\bf{W}}_B}} \right) + {\rm{Tr}}\left( {{\bf{QH}}_{B,n}^{}} \right) + 1} \right) - {\rm{Tr}}\left( {{\bf{H}}_{B,n}^{}{{\bf{W}}_B} + {\bf{QH}}_{B,n}^{} + {\bf{H}}_{D,n}^{}{\bf{W}}} \right) - 1} \right)} }\\
 + {\mathop{\rm Tr}\nolimits} \left( {{\bf{A}}\left( {1 - {{\bf{e}}^{{\beta ^{worst*}}}}} \right)\left( {{\bf{G}}_B^H{{\bf{W}}_B}{\bf{G}}_B^{} + {\bf{G}}_B^H{\bf{QG}}_B^{} + {{\bf{I}}_L}} \right) + {\bf{AG}}_E^H{\bf{WG}}_E^{}} \right) - {\mathop{\rm Tr}\nolimits} ({\bf{BW}}) - {\mathop{\rm Tr}\nolimits} ({\bf{C}}{{\bf{W}}_B}) - {\mathop{\rm Tr}\nolimits} ({\bf{DQ}}).
\end{array}
 \end{equation}
\end{small}
\end{figure*}

\begin{appendices}
  \section{Proof of Proposition 1}
Let ${\beta ^{worst}}_{}^ * $ be the optimal solution of the outer-level optimization problem, ${\bf{W}}_{}^ * ,{\rm{ }}{\bf{W}}_B^ * ,{\rm{ }}{\bf{Q}}_{}^ *$ and $ \varphi _{}^{worst * }$ be the optimal solution of the inner-level problem (45) with fixed ${\beta ^{worst}}_{}^ * $ (in this case, ${\bf{W}}_{}^ * ,{\rm{ }}{\bf{W}}_B^ * ,{\rm{ }}{\bf{Q}}_{}^ *$ and $\varphi _{}^{worst * }$ are also the optimal solution of (44)). Then, we formulate the following power minimization problem (54) on the top of this page. Let ${\bf{\bar W}},{\rm{ }}{\bf{\bar W}}_B^{}$ and ${\bf{\bar Q}}$ be the optimal solutions of (54), following Appendix B in \cite{6482662}, we have ${\bf{\bar W}},{\rm{ }}{\bf{\bar W}}_B^{}$ and ${\bf{\bar Q}}$ are also the optimal solution of (45) with fixed ${\beta ^{worst*}}$.

Then, we write the Lagrangian associated with (54) as (55) on the top of previous page, in which ${\bf{A}} \succeq {\bf{0}},{\bf{B}} \succeq {\bf{0}},{\bf{C}} \succeq {\bf{0}},{\bf{D}} \succeq {\bf{0}},\varepsilon_{m}  \ge 0,\lambda  \ge 0,\mu  \ge 0$ and ${\nu _n} \ge 0,{\rm{ }}\forall n \in {{\cal N}}$ are the dual variables corresponding to primal constraints in (54). Let ${{\bf{I}}_\varepsilon }{\rm{ = diag([}}{\varepsilon _1}{\rm{,}}{\varepsilon _2}, \ldots ,{\varepsilon _M}{\rm{])}}$, the KKT conditions that are relevant to  ${\bf{\bar W}}$ are given by

 \begin{equation}
{\bf{I}} + {{\bf{I}}_\varepsilon } + \mu \left( {{{\bf{e}}^{R_C^{worst*}}} - 1} \right){\bf{H}}_C^{} - \sum\limits_{n = 1}^N {{\nu _n}{\bf{H}}_{D,n}^{}}  + {\bf{G}}_E^{}{\bf{AG}}_E^H - {\bf{B}} = \bf{0}
 \end{equation}
 \begin{equation}
\begin{array}{l}
{\nu _n}\left( {} \right.{{\bf{e}}^{\varphi _{}^{worst*}}}{\mathop{\rm Tr}\nolimits} \left( {{\bf{H}}_{B,n}^{}{{\bf{W}}_B}} \right) + {{\bf{e}}^{\varphi _{}^{worst*}}}{\mathop{\rm Tr}\nolimits} \left( {{\bf{QH}}_{B,n}^{}} \right) + {{\bf{e}}^{\varphi _{}^{worst*}}}\\
 - {\mathop{\rm Tr}\nolimits} \left( {{\bf{H}}_{B,n}^{}{{\bf{W}}_B}} \right) - {\mathop{\rm Tr}\nolimits} \left( {{\bf{QH}}_{B,n}^{}} \right) - {\mathop{\rm Tr}\nolimits} \left( {{\bf{H}}_{D,n}^{}{\bf{W}}} \right) - 1\left. {} \right){\rm{ = }}0
\end{array}
 \end{equation}
 \begin{equation}
{\bf{\bar W}} \succeq {\bf{0}}
 \end{equation}
 \begin{equation}
{\bf{B\bar W}}{\rm{ = }}{\bf{0}}.
 \end{equation}
Postmultiplying (56) by ${\bf{\bar W}}$ and making use of (59) yield
 \begin{equation}
\left( {{\bf{I}} +{{\bf{I}}_\varepsilon } + \mu \left( {{{\bf{e}}^{R_C^{worst*}}} - 1} \right){\bf{H}}_C^{}{\rm{ + }}{\bf{G}}_E^{}{\bf{AG}}_E^H} \right){\bf{\bar W}}= \left( {\sum\limits_{n = 1}^N {{\nu _n}{\bf{H}}_{D,n}^{}} } \right){\bf{\bar W}}.
 \end{equation}
Since ${\bf{I}} + {{\bf{I}}_\varepsilon } + \mu \left( {{{\bf{e}}^{R_C^{worst*}}} - 1} \right){\bf{H}}_C^{}{\rm{ + }}{\bf{G}}_E^{}{\bf{AG}}_E^H \succ 0$, we have
\begin{equation}
\begin{array}{l}
{\rm{Rank}}\left( {{\bf{\bar W}}} \right){\rm{ = Rank}}\left( {\left( {\sum\limits_{n = 1}^N {{\nu _n}{\bf{H}}_{D,n}^{}} } \right){\bf{\bar W}}} \right)\le \min \left( {{\rm{Rank}}\left( {\sum\limits_{n = 1}^N {{\nu _n}{\bf{H}}_{D,n}^{}} } \right),{\rm{Rank}}\left( {{\bf{\bar W}}} \right)} \right).
\end{array}
  \end{equation}
From (61) we still can hardly calculate the rank of ${\bf{\bar W}}$, since the rank of $\sum\limits_{n = 1}^N {{\nu _n}{\bf{H}}_{D,n}^{}} $ is hardly to be known. However, combining complementary relaxation condition (57) and ${\rm{C6'''}}$in (54), we can conclude that for any $\forall n \in {{\cal N}}$
\begin{equation}
 \left\{ \begin{array}{l}
{\nu _n}\! \ge\! 0,\!{\rm{if }}\!\log \!\left(\! {1\! +\! \frac{{{\mathop{\rm Tr}\nolimits} \left( \!{{\bf{H}}_{D,n}^{}{\bf{W}}} \!\right)}}{{{\mathop{\rm Tr}\nolimits} \left(\! {{\bf{H}}_{B,n}^{}{{\bf{W}}_B}} \!\right)\! + \!{\mathop{\rm Tr}\nolimits} \left(\! {{\bf{QH}}_{B,n}^{}} \!\right)\! + 1}}} \!\right) \!-\! \varphi _{}^{worst*} \! =\! 0\\
{\nu _n}\! = \!0,\!{\rm{if}}\log \!\left(\! {1\! +\! \frac{{{\mathop{\rm Tr}\nolimits} \left( \!{{\bf{H}}_{D,n}^{}{\bf{W}}} \!\right)}}{{{\mathop{\rm Tr}\nolimits} \left( \!{{\bf{H}}_{B,n}^{}{{\bf{W}}_B}}\! \right)\! + \!{\mathop{\rm Tr}\nolimits} \left( \! {{\bf{QH}}_{B,n}^{}}\! \right)\! + \!1}}}\! \right)\! -\! \varphi _{}^{worst*} \! >\! 0.
\end{array} \right.
   \end{equation}
 Hence, if there has only one $\tilde n$ that makes $\log \left( {1 + \frac{{{\mathop{\rm Tr}\nolimits} \left( {{\bf{H}}_{D,\tilde n}^{}{\bf{W}}} \right)}}{{{\mathop{\rm Tr}\nolimits} \left( {{\bf{H}}_{B,\tilde n}^{}{{\bf{W}}_B}} \right) + {\mathop{\rm Tr}\nolimits} \left( {{\bf{QH}}_{B,\tilde n}^{}} \right) + 1}}} \right) - \varphi _{}^{worst*} {\rm{ = }}0$, we have $\sum\limits_{n = 1}^N {{\nu _n}{\bf{H}}_{D,n}^{}}  \to {\nu _{\tilde n}}{\bf{H}}_{D,\tilde n}^{}{\rm{\; or\; }}{\bf{0}}$ which is of rank one and then we can drive the following inequation (63) from (61), which implies ${\rm{Rank}}\left( {{\bf{\bar W}}} \right){\rm{ = }}\;1\; {\rm{ or\; 0}}$. Because ${\rm{Rank}}\left( {{\bf{\bar W}}} \right){\rm{ = 0}}$ means all zero matrix which is meaningless, we have ${\rm{Rank}}\left( {{\bf{\bar W}}} \right){\rm{ = }}1$.

\begin{equation}
\begin{array}{l}
{\rm{Rank}}\left( {{\bf{\bar W}}} \right) \le \min \left( {{\rm{Rank}}\left( {{\nu _{\tilde n}}{\bf{H}}_{D,\tilde n}^{}} \right){\rm{ \;or\;0,Rank}}\left( {{\bf{\bar W}}} \right)} \right)\min \left( {1\;{\rm{ or \;0}},{\rm{Rank}}\left( {{\bf{\bar W}}} \right)} \right).
\end{array}
 \end{equation}

 \end{appendices}

\small
\bibliographystyle{IEEEtran}
\bibliography{ref}

\end{document}